\documentclass[conference]{IEEEtran}
\IEEEoverridecommandlockouts

\usepackage{cite}
\usepackage{amsmath,amssymb,amsfonts}
\usepackage{graphicx}
\usepackage{textcomp}
\usepackage{xcolor}
\usepackage{fancyhdr}
\usepackage[hyphens]{url}
\usepackage[normalem]{ulem}
\usepackage{outlines}
\usepackage{algpseudocode,algorithm}
\usepackage{physics}
\usepackage{multirow}
\usepackage{booktabs}
\usepackage[caption=true]{subfig}
\usepackage[export]{adjustbox}
\usepackage{comment}
\usepackage{xcolor}
\usepackage{hyperref}

\newcommand{\nuop} {{NuOp}}
\newcommand{\cirq} {{Cirq}}

\begin{document}

\title{Designing Calibration and Expressivity-Efficient Instruction Sets for Quantum Computing\\
\thanks{*Prakash Murali and Lingling Lao are joint first authors. Corresponding author: pmurali@cs.princeton.edu}
}
\author{\IEEEauthorblockN{Prakash Murali$^*$}
    \IEEEauthorblockA{\textit{Princeton University}}
\and
    \IEEEauthorblockN{Lingling Lao$^*$}
    \IEEEauthorblockA{\textit{University College London}}
\and
    \IEEEauthorblockN{Margaret Martonosi}
    \IEEEauthorblockA{\textit{Princeton University}}
\and
    \IEEEauthorblockN{Dan Browne}
    \IEEEauthorblockA{\textit{University College London}}
\and
}

\maketitle
\thispagestyle{plain}
\pagestyle{plain}
\begin{abstract}
Near-term quantum computing (QC) systems have limited qubit counts, high gate (instruction) error rates, and typically support a minimal instruction set having one type of two-qubit gate (2Q).
To reduce program instruction counts and improve application expressivity, vendors have proposed, and shown proof-of-concept demonstrations of richer instruction sets such as XY gates (Rigetti) and fSim gates (Google). 
These instruction sets comprise of families of 2Q gate types parameterized by continuous qubit rotation angles. That is, it allows a large set of different physical operations to be realized on the qubits, based on the input angles.
However, having such a large number of gate types is problematic because each gate type has to be calibrated periodically, across the full system, to obtain high fidelity implementations. This results in substantial recurring calibration overheads even on current systems which use only a few gate types.
Our work aims to navigate this tradeoff between application expressivity and calibration overhead, and identify what instructions vendors should implement to get the best expressivity with acceptable calibration time. 

Studying this tradeoff is challenging because of the diversity in QC application requirements, the need to optimize applications for widely different hardware gate types and noise variations across gate types. Therefore, our work develops \nuop, a flexible compilation pass based on numerical optimization, to efficiently decompose application operations into arbitrary hardware gate types. Using \nuop\ and four important quantum applications, 
we study the instruction set proposals of Rigetti and Google, with realistic noise simulations and a calibration model. Our experiments show that implementing 4-8 types of 2Q gates is sufficient to attain nearly the same expressivity as a full continuous gate family, while reducing the calibration overhead by two orders of magnitude. With several vendors proposing rich gate families as means to higher fidelity, our work has potential to provide valuable instruction set design guidance for near-term QC systems.  
\end{abstract}


\section{Introduction}

\begin{table*}[]
\centering
\begin{tabular}{|c|c|c|c|}
\hline
\multicolumn{2}{|c|}{\textbf{Rigetti}} & \multicolumn{2}{c|}{\textbf{Google}} \\ \hline
Current          & Anticipated         & Current         & Anticipated        \\ \hline
\begin{tabular}{@{}c@{}} $\resizebox{.15\hsize}{!}{$
                    \mathrm{CZ}=\begin{pmatrix}
                    1 & 0 & 0 & 0\\ 
                    0 & 1 & 0 & 0\\ 
                    0 & 0 & 1 & 0\\ 
                    0 & 0 & 0 & -1
                    \end{pmatrix}$}$
                \\ XY$(\pi)$\end{tabular}
                & \begin{tabular}{@{}c@{}}
                    $\resizebox{.25\hsize}{!}{$
                    \mathrm{XY}(\theta)=\begin{pmatrix}
                    1 & 0 & 0 & 0\\ 
                    0 & \mathrm{cos}(\theta/2) & i\mathrm{sin}(\theta/2) & 0\\ 
                    0 & i\mathrm{sin}(\theta/2) & \mathrm{cos}(\theta/2) & 0\\ 
                    0 & 0 & 0 & 1
                    \end{pmatrix}$}$\\ CZ \end{tabular}
                    & 
                    
                    \begin{tabular}{@{}c@{}} SYC = fSim$(\pi/2, \pi/6)$  \\
                             %   iSWAP = fSim$(\pi/2, 0)$ \\ 
                             $\sqrt{i\mathrm{SWAP}}=\mathrm{fSim}(\pi/4, 0)$ 
                             %   \\ $\mathrm{SWAP}\sim \mathrm{fSim}(\pi/2, \pi)$ 
                               \end{tabular}

                    &    \resizebox{.3\hsize}{!}{$
                    \mathrm{fSim}(\theta, \phi)=\begin{pmatrix}
                    1 & 0 & 0 & 0\\ 
                    0 & \mathrm{cos}(\theta) & -i\mathrm{sin}(\theta) & 0\\ 
                    0 & -i\mathrm{sin}(\theta) & \mathrm{cos}(\theta) & 0\\ 
                    0 & 0 & 0 & e^{-i\phi}
                    \end{pmatrix}$.}
                    \\ \hline
Fidelity $\sim$95\%                 & 95-99\%                    & $\sim$99.6\%                 & $\sim$99.6\%                    \\ \hline
\end{tabular}
\caption{Current and anticipated two-qubit gate types in Rigetti and Google systems. Each two-qubit gate type corresponds to a 4x4 unitary matrix. Current gate sets are based on Rigetti's Aspen-8 \cite{rigetti_aspen8} and Google's Sycamore \cite{arute2019quantum} devices. Anticipated gate sets are based on \cite{abrams2019implementation} and \cite{google_continuous_gate}. Controlled Z (CZ) is an example of a fixed gate type. In contrast, fSim($\theta, \phi$) is parameterized by $\theta, \phi$. Varying these parameters allows a broad range of gate types to be realized in hardware. XY($\theta$) is a subset of fSim gate family up to single-qubit rotations. (The actual instruction sets also include single-qubit operations for universality, they are not shown here.)
}
\label{fig:systems_gate_sets}
\end{table*}

Quantum computing (QC) is an emerging paradigm that uses quantum mechanical principles to manipulate information. QC systems store information using \emph{qubits} (quantum bits) and manipulate information using \emph{gates} (operations). These operations allow the system to exploit quantum effects such as superposition, entanglement and interference to explore large state spaces efficiently, sometimes faster than classical (non-quantum) systems. In the near-future, QC is poised to drive research in domains such as machine learning \cite{quantum_ml1}, quantum chemistry \cite{vqe1} and material science \cite{quantum_matscience}. Several prototype QC systems have been built using hardware technologies such as superconducting and trapped-ion qubits. These systems are referred to as Noisy Intermediate-Scale Quantum (NISQ) systems \cite{nisq} and have small qubit counts, restricted connectivity and high gate error rates. 
Although resource-constrained, NISQ systems are useful for demonstrating near-term applications \cite{nam2019ground} and milestones such as quantum supremacy \cite{arute2019quantum}. 

Typically, NISQ systems support a \emph{universal} gate set composed of continuous single-qubit (1Q) rotations and a few two-qubit (2Q) gate types that can express any application operation. Among them, 2Q gates generate entanglement between pairs of qubits and constitute a key building block for quantum algorithms. However, 2Q gates are more challenging to implement than 1Q gates and have higher error rates. For example, on IBM systems, 2Q gate error rates are 1-5\% and 1Q gate error rates are less than 0.1\%. Since a single entangling 2Q gate is sufficient for universality, most vendors use instruction sets having only one 2Q gate type. For example, in Rigetti's and Google's early QC systems, Controlled Z (CZ) gate was the only supported 2Q gate type \cite{rigetti_19q, google_9q, Boixo2018}. 
% \begin{equation}
%         \label{eqn:cz}
%         \resizebox{.35\hsize}{!}{$
%         \mathrm{CZ}=\begin{pmatrix}
%         1 & 0 & 0 & 0\\ 
%         0 & 1 & 0 & 0\\ 
%         0 & 0 & 1 & 0\\ 
%         0 & 0 & 0 & -1\\
%         \end{pmatrix}$}
% \end{equation}

The type and error rate of the hardware 2Q gate determines \emph{application expressivity}, that is, the number of instructions required to implement an application and the overall execution fidelity.
%Figure \ref{} shows decompositions where application operations have different instructions counts for two hardware gate types. \todo{}
When a system has a single 2Q gate type, the compiler is forced to express all application operations using it, worsening executable instruction counts and duration. Importantly, the error rate of the available 2Q gate constrains the number of program instructions that can be reliably executed. Although vendors have been improving gate implementations, progress has been slow because of challenges in accurate qubit control, qubit defects from lithographic manufacturing and emergent sources of noise. Therefore, if an application cannot be expressed concisely, it is unlikely to execute successfully on near-term hardware. 
%has a low likelihood of a successful run.

To improve expressivity, and thereby improve execution success rate, industry and academic vendors~\cite{abrams2019implementation, google_continuous_gate, eth_two_qubit_gate} have proposed instruction sets which have continuous 2Q gate families. Table \ref{fig:systems_gate_sets} shows the proposed XY gate family from Rigetti and the fSim gate family from Google.
% such as the CPHASE (CZ) gates from ETH~\cite{}.
 Each gate family is parameterized by rotation angles (e.g., $\theta$ and $\phi$ for Google's fSim gates) and forms a continuous set of two-qubit gate types (different unitaries). This means, in theory, there are an infinite number of hardware 2Q gate types that can be used to express applications succinctly. In practice, however, quantum gates are implemented using analog pulse sequences which must be calibrated to obtain high fidelity. Calibration routines require a large number of test executions to tune control parameters and characterize the hardware operations.
% and fine-tune their pulse sequences.
Worse, calibration is not a one-time step (parameters drift over time) and must be performed periodically to maintain high fidelity. Even with a single type of 2Q gate, Google's 54-qubit device requires up to four hours of calibration per day \cite{arute2019quantum}. Therefore, across the full parameter space of a continuous gate set, calibration overheads are likely enormous, especially as devices scale up.

Our work studies the tradeoff between application expressivity and calibration overhead to determine what gate sets and gate parameter combinations vendors should implement to offer high-fidelity executions without tremendous calibration time. To accurately compare instruction set options, we require a toolflow to automatically decompose and optimize applications for different candidate gate sets. This is challenging because of the variety of gate options, variations in fidelity across gate parameter values, and the need to use different optimization strategies based on the hardware characteristics. 

\begin{figure}[t]
    \centering
    \includegraphics[width=0.50\textwidth]{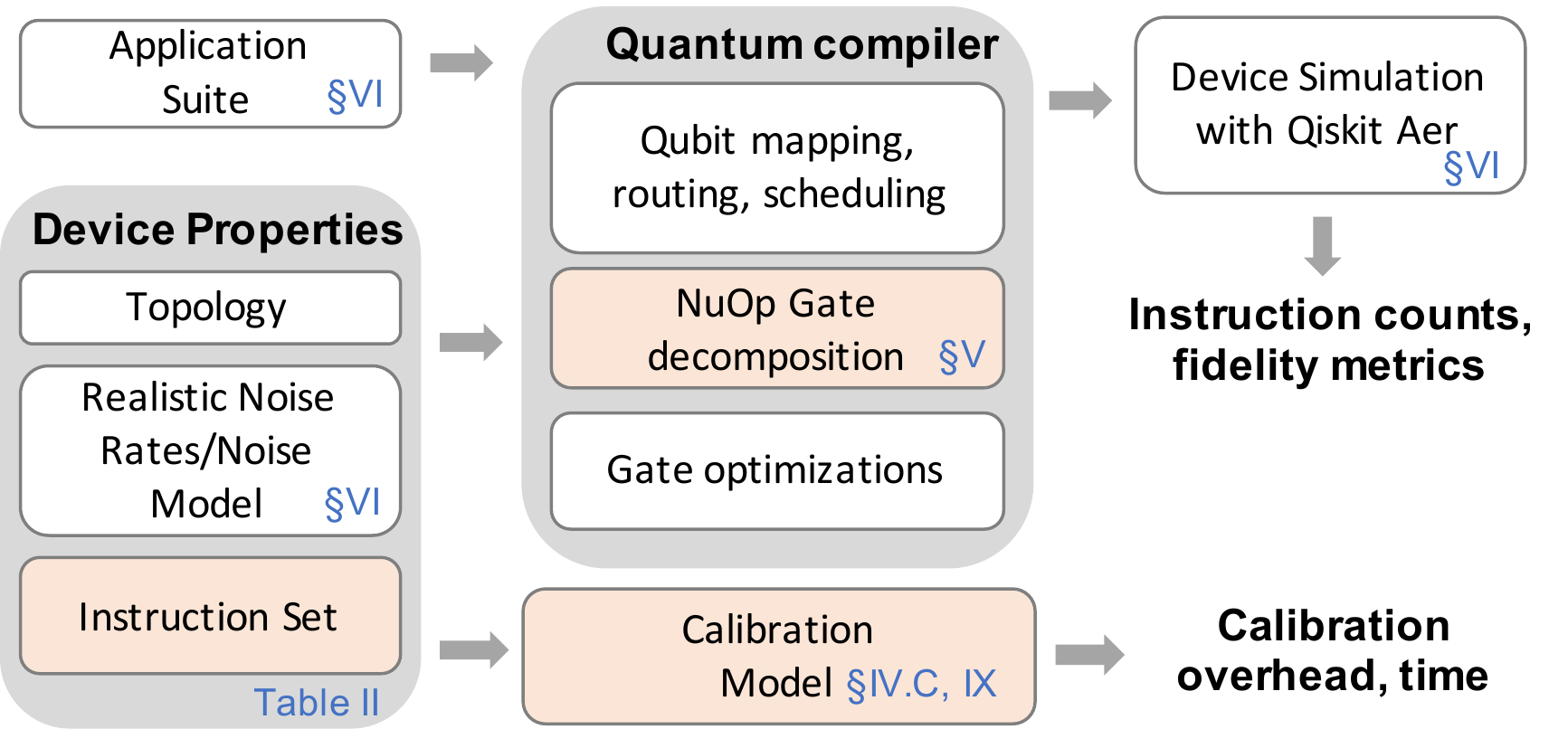}
    \caption{Simulation framework. To study different instruction set choices, our work performs an application-driven simulation study with realistic device and calibration models.}
    \label{fig:framework}
\end{figure}

Figure \ref{fig:framework} shows our simulation framework. Our work addresses the toolflow challenge by building \nuop, a flexible compilation pass based on numerical optimization techniques. \nuop\ accepts as input a set of target hardware gate types and uses noise-adaptive compilation across gate parameter values and approximate gate decomposition to heavily optimize applications, enabling accurate comparisons across gate choices. Using \nuop, we perform a simulation-driven study to identify the best instruction design choices across two vendor gate families and several NISQ applications. 
%Using our pass, we perform a simulation-driven study to identify the best gate choices across two vendor gate sets and several NISQ applications.

Our contributions include:
\begin{itemize}
\item Current QC instruction sets belong to one of two camps: minimal sets having a single two-qubit gate type \cite{ibm_devices} (poor expressivity, easy to calibrate) or families of continuous gate sets \cite{google_continuous_gate, abrams2019implementation, eth_two_qubit_gate} (rich expressivity, hard to calibrate). These two choices and their calibration tradeoffs are reminiscent of the tradeoffs in classical RISC vs. CISC ISA choices. Our work is the first to study the expressivity vs. calibration tradeoff in QC, and design an instruction set that has rich expressivity and is easy to calibrate.

\item To design this instruction set, we use \nuop~to characterize instruction counts for four representative QC applications across the continuous parameter space for XY and fSim gate families. Across the parameter space, one to six hardware 2Q gates are required to implement an application 2Q operation.
%Instruction counts per application operation vary from one to six hardware gates, across the gate parameter space. 
Based on this characterization, we identify a small subset (4-8 discrete combinations) of these gate families that offers near-optimal expressivity i.e., instruction counts and fidelity that are better than instruction sets with a single gate type and close to the optimal values that can be obtained using the entire continuous parameter space of XY and fSim families.

\item  Using a realistic calibration model for fSim gates, we show that the proposed gate sets having 4-8 gate types, save two orders of magnitude calibration overhead, compared to using the entire parameter space. Since calibration is not a one-time overhead, this large reduction makes our instruction set practically viable for real-system implementations.

\item Finally, QC systems are known to have highly variable error rates across qubits and time~\cite{murali2019noise, triq_isca, tannu2019not, nishio2020extracting}. We demonstrate that exposing multiple 2Q gate types is a viable method to mitigate such noise variations. When a noise-adaptive compiler pass such as \nuop~is used in conjunction with multiple gate types, the compiler can choose different gate types on different qubits to obtain significant improvements in fidelity.

\end{itemize}

Our work makes the case for an instruction set having a small number of expressive two-qubit gate types. With major QC vendors aiming to offer more instruction types in the near-term \cite{google_continuous_gate, eth_two_qubit_gate}, we expect our work to significantly impact industry and academic instruction sets and provide useful guidance to hardware designers.

\section{Background on Quantum Gates}

{\noindent \textbf{Gate family vs. type:}} Quantum algorithms encode information using a set of qubits and  manipulate information by applying different quantum gates (instructions).
Each $n$-qubit quantum gate can be uniquely defined by a unitary matrix $U_{2^{n}\times2^{n}}$.
%For example, the \textit{fixed} unitary of the CZ gate is shown in Table \ref{fig:systems_gate_sets}.
The unitary matrix of a gate may be a \emph{fixed unitary} or a \emph{parameterized unitary}. An example of a fixed unitary gate is the CZ gate shown in Table \ref{fig:systems_gate_sets}.
In contrast, fSim($\theta, \phi$) is parameterized by $\theta$ and $\phi$. For such gates, we use the term \emph{gate family} since it allows an infinite number of unitaries to be realized based on the parameter values. We use the term \emph{gate type} to refer to fixed parameter values in this family e.g., fSim$(\pi, \pi/2)$ and fSim$(\pi/6, \pi/8)$ and CZ = fSim$(0, \pi$) are three distinct gate types. 

{\noindent \textbf{Gate Implementation in NISQ systems:}}
% Time-dependent analog control pulses are applied to one or more qubits to realize a desired state transition (gate). 
Gate implementation depends on the qubit hardware technology and gate type. For example, in Google's Sycamore system (transmon qubits), single-qubit gates are implemented using microwave pulses that are resonant with the qubit frequency; two-qubit gates (SYC and $\sqrt{\mathrm{iSWAP}}$) are implemented by bringing neighboring qubits in resonance through a tunable coupler that mediates the interaction \cite{arute2019quantum}. 
Gate implementations in all NISQ systems are error-prone. The quality of a gate implementation is characterized by its fidelity ($1 -$error rate). Table \ref{fig:systems_gate_sets} shows average gate fidelities for Rigetti and Google systems.

\section{Related Work}

Most prior works~\cite{quil_isa,openqasm,eqasm, 9076341,2001.08838,8972540} consider instruction set design issues assuming that systems have only one two-qubit gate type. For example, Rigetti Quil \cite{quil_isa} and IBM OpenQASM \cite{openqasm} target only the CZ and CNOT gates, respectively. 
%eQASM \cite{eqasm, 9076341} is an academic programming interface which focuses on a VLIW QC architecture.  
Murali et al. \cite{triq_isca} compare native and software-visible instruction sets and recommend exposing all available gate types to the compiler to improve application fidelity. 
In contrast to these minimal instruction sets, we focus on instruction sets with multiple gate types to improve application expressivity. 
%continuous 2Q gate families proposed for Google and Rigetti systems and explores the best way to design 2Q instruction sets.

Google's Sycamore device has preliminary support for a continuous fSim gate \cite{google_continuous_gate}. Rigetti's XY gates were demonstrated in \cite{abrams2019implementation}. 
Lacroix et al. \cite{eth_two_qubit_gate} experimentally demonstrated that using continuous Controlled-Phase gates (CZ$(\theta$)) is beneficial for quantum optimization (QAOA). I
%t has also been theoretically proven that a continuous sets of XY gates \cite{peterson2020two} can provide significant reduction in application instruction counts. 
These works represent one extreme of the design space, i.e., the full parameter space is available to programs, which optimizes for application expressivity but requires demanding calibration schemes. Our work uses realistic gate errors and calibration models from Google and Rigetti devices to study what subset of the gate parameter space should be exposed to obtain high application reliability with low calibration overhead.

% The authors of \cite{eth_two_qubit_gate} propose and implement continuous controlled phased CPHASE$(\theta$) gates on a seven qubit superconducting chip. 

Our work requires an automated method to decompose and optimize applications for arbitrary vendor gate sets. Several works have investigated compilation and program optimization techniques for QC devices. Early works focused on minimizing metrics such as gate count and circuit depth \cite{mapping1, mapping2, mapping3, mapping4, li2019tackling}. Recently noise-adaptive compilation passes \cite{ murali2019noise,tannu2019not,nishio2020extracting, murali2020software} have been developed to adapt program executions to spatio-temporal fidelity variations in real systems. However, these works and compilers such as IBM Qiskit \cite{qiskit} and TriQ \cite{triq_isca} assume that the hardware instruction set has a single type of two-qubit gate. Hence, they are not directly applicable for studying instruction set design issues for multiple two-qubit gate types. 

\section{Instruction Set Design Choices and Tradeoffs}
\label{sec:instruction_set}
\begin{figure*}[t]
    \centering
    \subfloat[Two-qubit QV unitary]{
    \includegraphics[scale=0.60]{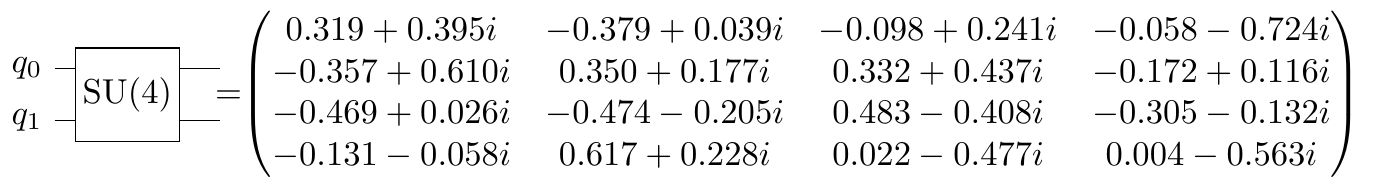}
    \label{fig:qv_u1}
    }
     \subfloat[Two-qubit QAOA unitary]{
    \includegraphics[scale=0.60]{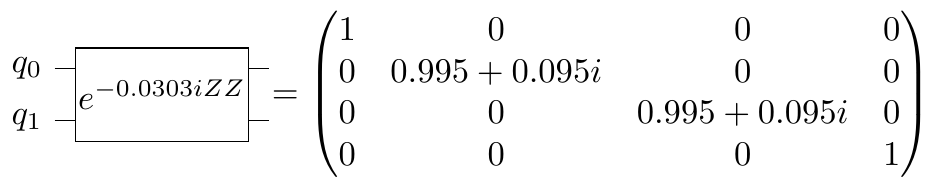}
    \label{fig:qaoa_u1}
    }
  
     \subfloat[Decomposition: QV unitary with CZ]{
        \includegraphics[scale=0.52]{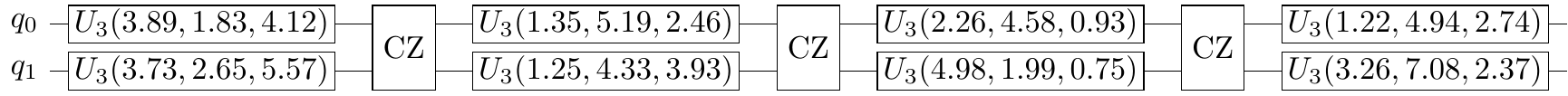}
        \label{fig:qv_u1_cz}
    }
      \subfloat[Decomposition: QAOA unitary with CZ]{
        \includegraphics[scale=0.50]{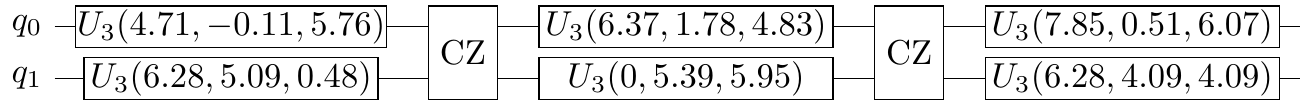}
        \label{fig:qaoa_u1_cz}
    }
    
    \subfloat[Decomposition: QV unitary with $\sqrt{\mathrm{iSWAP}}$]{
        \includegraphics[scale=0.50]{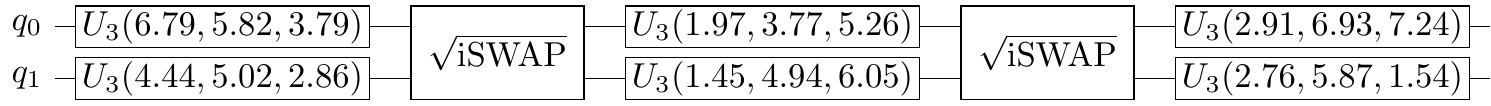}
        \label{fig:qv_u1_iSWAP}
    } 
    \subfloat[Decomposition: QAOA unitary with $\sqrt{\mathrm{iSWAP}}$]{
        \includegraphics[scale=0.50]{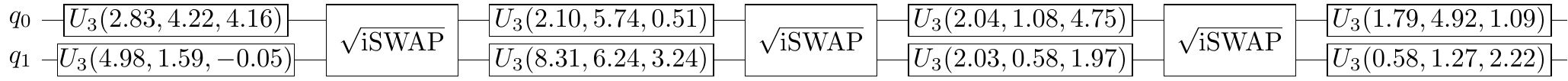}
        \label{fig:qaoa_u1_iSWAP}
    } 
    
    \caption{Decomposition examples using NuOp. (a) shows a two-qubit gate with unitary from the $\mathrm{SU(4)}$ group (Quantum volume circuits randomly sample gates from this group~\cite{cross2019validating}). 
    (b) shows a two-qubit Pauli interaction ($e^{-i\beta(Z\otimes Z)}$), which is used in QAOA circuits. 
    (c-f) show exact implementation (decomposition error $\approx 10^{-8}$) of these unitaries using different two-qubit hardware gates from Rigetti (CZ) and Google ($\sqrt{\mathrm{iSWAP}}$).
    The CZ gate is more expressive for QAOA unitaries and the $\sqrt{\mathrm{iSWAP}}$ gate is more expressive for QV unitaries. In (c-f), U3 gates are arbitrary single-qubit rotation operations\protect\footnotemark .}   
    \label{fig:decomposition_examples}
\end{figure*}
\begin{figure}[t]
    \centering
    \includegraphics[scale=0.27]{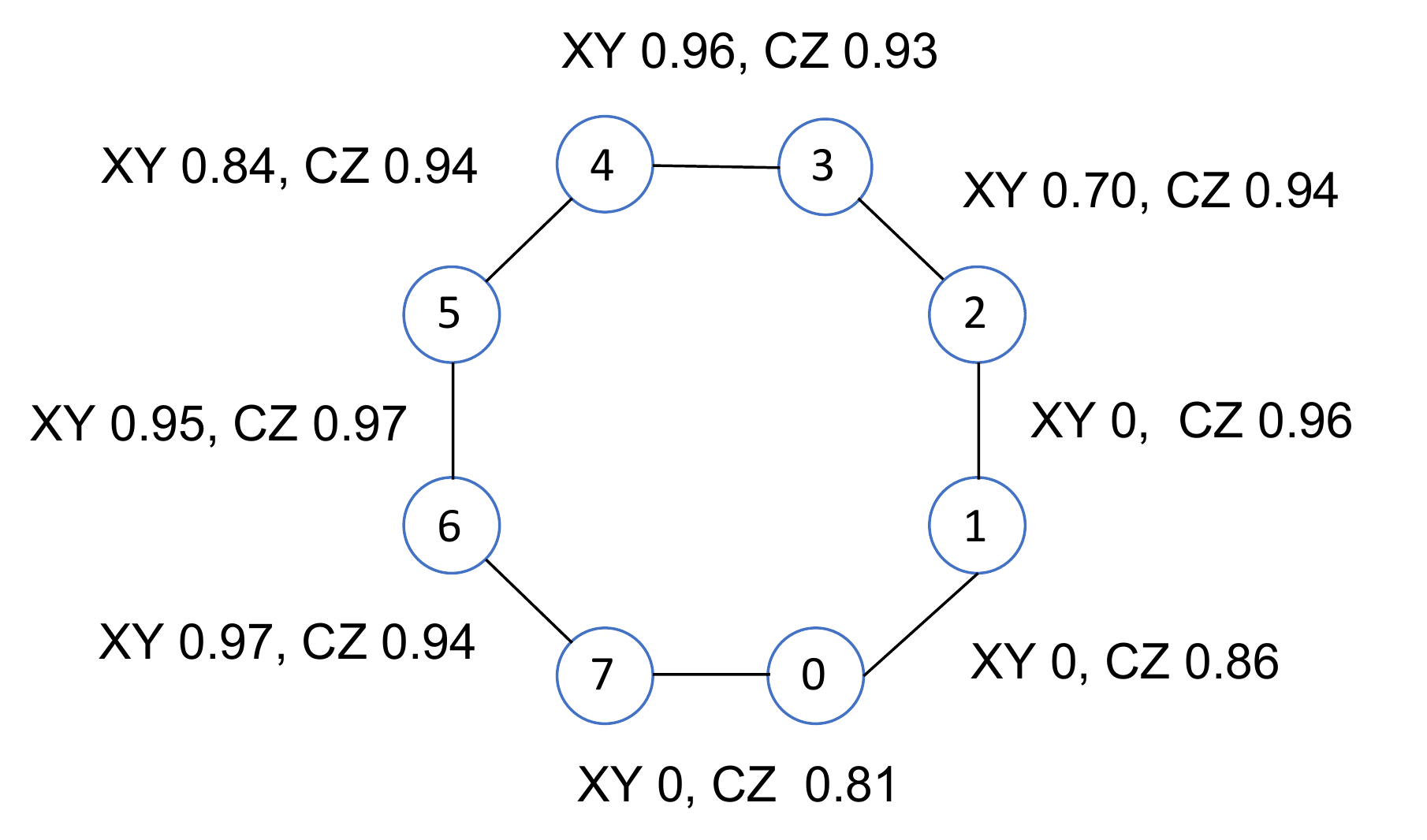}
    \caption{Layout of first 8 qubits in Rigetti Aspen-8 system. Nodes are qubits and edges are two-qubit gates labeled by the measured fidelity for XY$(\pi)$ and CZ gates \cite{rigetti_qcs}. The best gate varies across qubit-pairs.
    % and across time (not shown).
    }
    \label{fig:rigetti_lattice}
\end{figure}
\subsection{Background on ISAs with Continuous Two-Qubit Gates}

Rigetti has proposed using XY$(\theta)$ gates along with the Controlled-Z gate (CZ) as the instruction set in their future systems. An early demonstration of this gate set on a two-qubit system in \cite{abrams2019implementation} shows executions of broad range of gate types with fidelities ranging from 95\% to 99\%. Rigetti's current system (Aspen-8), a 30-qubit device, includes one gate type from this family, the $\mathrm{XY}(\pi)$ gate, in addition to the CZ gate. 

Similarly, Google has proposed a continuous fSim$(\theta, \phi)$ gate family for their devices. This gate family can be viewed as a composite of two families i.e., the CPHASE family, CZ($\phi$), obtained by varying $\phi$, fixing $\theta=0$, and the iSWAP-like family, iSWAP($\theta$), obtained by varying $\theta$, fixing $\phi=0$. 
%The $\mathrm{fSim}(\theta, \phi)$ gate can be seen as the composite of CPHASE gate (Equation~\ref{eqn:cz}, $\mathrm{CZ}(\phi)=\mathrm{fSim}(0, \phi)$) and iSWAP-like gate ($\mathrm{iSWAP}(\theta) = \mathrm{fSim}(\theta, 0) = \mathrm{XY}(0,2\theta)$). 
%
In \cite{google_continuous_gate}, Google demonstrated 525 instances of the fSim family with different $\theta$ and $\phi$, on two qubits. Recent experiments from their 54-qubit Sycamore processor use two gate types from this family: SYC gate = fSim$(\pi/2, \pi/6)$~\cite{arute2019quantum} and $\sqrt{\mathrm{iSWAP}}$ gate = fSim$(\pi/4, 0)$~\cite{arute2020quantum}, with fidelities as high as 99.4\%

In both Rigetti and Google instruction sets, multiple types of single-qubit gates (qubit rotations) are included for obtaining a universal gate set. However, these gates typically have one order of magnitude higher fidelity than two-qubit gates and have been optimized heavily using both hardware \cite{PhysRevA.96.022330} and software techniques \cite{qiskit}. Hence, our focus is on designing efficient two-qubit instruction sets.
%Single-qubit gates (qubit rotations) are easier to implement than two-qubit gates and have much higher fidelity. 

\subsection{Application-level Benefits of Multiple Gate Types}
Different gate types are advantageous to different QC applications.
For example, the CZ gates are useful in quantum error correction.
%and CNOT 
Excitation-preserving operations like the ones in XY and fSim families allow natural representations for quantum chemistry~\cite{vqe1,vqe2} and combinatorial optimization~\cite{farhi2014quantum}. Therefore, by offering more than one gate type, vendors aim to cater to the different application operation needs, and reduce the number of instructions required to express them.  

Figure \ref{fig:decomposition_examples} shows decompositions of two application unitaries using two types of hardware gates. When only one of these gate types is available, the compiler has to decompose all application unitaries using it, potentially increasing the number of instructions. Since NISQ systems have high two-qubit gate error rates, an increase in instruction count translates to a significant reduction in fidelity. Continuous gate families proposed by Rigetti and Google leverage this concept to an extreme. That is, they aim to offer a very large number of two-qubit gate types to applications using one or more controllable gate parameters. For Rigetti's XY family, Peterson et al.~\cite{peterson2020two} theoretically proves that offering applications access to the full continuous parameter space, can give $\sim30\%$ two-qubit instruction count reduction for generic quantum circuits. Similarly, Foxen et al.~\cite{google_continuous_gate} demonstrates that a continuous set of fSim gates can provide a 3X gate reduction compared to decomposition with one gate type. 
\footnotetext{An arbitrary single-qubit gate can be described by a unitary matrix with three parameters
\begin{equation}
    \label{eqn:u3}
   \textrm{U3}(\alpha, \beta, \lambda)=\begin{pmatrix}
\cos(\frac{\alpha}{2}) & -e^{i\lambda}\sin(\frac{\alpha}{2})\\ 
 e^{i\beta}\sin(\frac{\alpha}{2}) & e^{i(\beta+\lambda)}\cos(\frac{\alpha}{2})
\end{pmatrix}.\notag
\end{equation}}.
\subsection{Experimental Difficulties of Multiple Gate Types}

{\noindent \textbf{Calibration overhead:}}
Implementing multiple two-qubit basis gates in the same device with high fidelity is experimentally difficult and requires demanding calibration strategies.  First, each two-qubit gate on each qubit pair needs to be calibrated independently to find its optimal control parameters, and electronics need to be carefully calibrated to avoid any bleeding between control pulses of different gates. These are known challenges for continuous gate sets~\cite{google_continuous_gate}. Second, on real-systems, calibration of different combinations of single-qubit and multi-qubit gates, and parallel operations are also required to obtain high fidelity~\cite{klimov2020snake}.
Calibration complexity increases at least linearly with the number of two-qubit basis gates and becomes enormous for a continuous gate set. Since calibration parameters drift over time (causing gate error rate fluctuations of up to 10X~\cite{google_continuous_gate}), frequent re-calibration is also required in practice. In Section \ref{sec:calibration}, we model fSim gate calibration.

{\noindent \textbf{Fidelity variations across gate types:}}
Continuous gate families also have variable fidelities across the parameter space, qubits in the system and time (similar to variability seen in systems with one type of two-qubit gate \cite{triq_isca, tannu2019not}). 
Figure \ref{fig:rigetti_lattice} shows that the best gate type varies across qubits in Rigetti's Aspen-8 system. For XY gates, gate error rates vary up to 4X depending on the angle $\theta$ (see Figure 6 in \cite{abrams2019implementation}). For fSim gates, the error rates vary up to 5X based on $\theta$ and $\phi$ (see Figure 4 in \cite{google_continuous_gate}). If applications use a single gate type across the system, it may result in severe fidelity losses. Instead, by using multiple gate types with noise-aware compilation, applications can benefit from the best gates on each part of the device.

\subsection{Our Work}
Our work studies key instruction design questions considering application-level benefits, calibration overheads and noise variability. \emph{Is it possible to design an instruction set having a small set of discrete gate types, but achieving the expressivity and fidelity of a continuous gate family? What are the calibration overheads of such an instruction set compared to a continuous gate family? Is it beneficial to expose multiple gate types and use noise adaptivity across gate types to mitigate noise variations?}  
%multiple gate types have noise variations, can we use noise-aware compilation across gate types or parameter values to improve application fidelity?

Table \ref{tab:instruction} shows the gate set options considered: instruction sets with a single type of two-qubit gate (S1-S7), our proposed instruction sets for Rigetti (R1-R5) and Google (G1-G7) and fully continuous Full$_{\mathrm{XY}}$ and Full$_{\mathrm{fSim}}$ gate families. To study these options, we use the framework shown in Figure \ref{fig:framework}. 
%The toolflow is configurable based on an input gate set and allows us to obtain optimized executables across applications and devices.

\begin{table*}[]
\footnotesize
\centering
\begin{tabular}{|l|l|}
\hline
\textbf{Type}                      & \textbf{Instruction Sets} (Up to single-qubit rotations, XY($\theta$) = iSWAP($\theta/2$) = fSim($\theta/2, 0$), CZ($\phi$) = fSim($0, \phi$).)                    \\ \hline
\textbf{S}ingle two-qubit gate type &
  \begin{tabular}[c]{@{}l@{}}
      S1 = SYC = fSim($\pi/2, \pi/6$), 
      S2 = $\sqrt{\mathrm{iSWAP}}$ = fSim($\pi/4, 0$), 
      S3 = CZ = fSim($0, \pi$), \\ 
      S4 = iSWAP = fSim($\pi/2, 0$), 
      S5 = fSim($\pi/3, 0$), 
      S6 = fSim($3\pi/8, 0$), 
      S7 = fSim($\pi/6, \pi$)
  \end{tabular} \\ \hline
Multiple gate types (\textbf{G}oogle) &
  \begin{tabular}[c]{@{}l@{}}G1 = \{S1,S2\} , G2 = \{S1,S2,S3\}, G3 = \{S1,S2,S3,S4\}, G4 = \{S1,S2,S3,S4,S5\}, G5 = \{S1,S2,S3,S4,S5,S6\}, \\ G6 = \{S1,S2,S3,S4,S5,S6,S7\}, G7 = \{S1,S2,S3,S4,S5,S6,S7,SWAP\}, \end{tabular} \\ \hline
Multiple gate types (\textbf{R}igetti) &
  \begin{tabular}[c]{@{}l@{}}R1 = \{S3,S4\}, R2 = \{S2,S3,S4\},  R3 = \{S2,S3,S4,S5\}, R4 = \{S2,S3,S4,S5,S6\}, R5 = \{S2,S3,S4,S5,S6,SWAP\}\end{tabular} \\ \hline
Full continuous gate set & Full$_{\mathrm{XY}}$ = \{XY($\theta$): $\theta \in [0, \pi]$\}, Full$_{\mathrm{fSim}}$ = \{fSim($\theta, \phi$): $\theta, \phi \in [0, \pi]$\} \\ \hline
\end{tabular}
\caption{Instruction sets studied in this work. Each set also includes arbitrary single qubit operations (not shown) to make it universal. S1-S7 were selected based on our application characterization experiments in Section \ref{sec:benefits}. See Figure \ref{fig:fsim_heatmap} for a graphical view of S1-S7 on the fSim parameter space. For Google, G1-G7 are combinations of S1-S7, e.g., G2 has three two-qubit gate types. For Rigetti, R1-R5 are selected based on subsets which can be supported using the XY gate family. (For ease of presentation, the table expresses Rigetti gates with fSim notation; they can be translated to XY($\theta$) notation using the identities on the top of the table.)}
\label{tab:instruction}
\end{table*}

\section{Toolflow for Architectural Exploration}
\label{sec:decomposition}
QC applications are typically written in terms of high-level device and gate set-independent operations. During compilation, program qubits are mapped onto the target device, communication operations such as SWAPs are inserted to comply with limited connectivity, and operations are translated or decomposed into the target instruction set. Figure \ref{fig:decomposition_examples} shows gate decomposition examples. Gate decomposition methods aim to represent the unitary of a desired application operation within a tolerable inaccuracy, using a circuit with short depth (to minimize decoherence), and small number of two-qubit gates (to minimize gate errors).

%To study different instruction choices using generic QC applications, we require a method to express and optimize these applications for a given gate set. 
To probe instruction set design issues across different gate types and their combinations we require a flexible decomposition method that can support any application operation and hardware gate type. Current industry compilers focus on gate decomposition routines for specific gate types of interest to a particular hardware platform. These compilers typically use linear algebra-based methods such as KAK decomposition ~\cite{tucci1999rudimentary,khaneja2001time,kraus2001optimal, vatan2004optimal,mottonen2004quantum, iten2017quantum} or hard coded gate identities \cite{maslov2005}. These methods are difficult to extend to continuous gate sets and arbitrary parameter combinations within these gate families.
For example, IBM's Qiskit compiler includes routines for decomposing arbitrary two-qubit unitaries to CNOT gates. Rigetti's Quilc \cite{quilc} includes support for CZ and XY$(\pi)$ gates, but does not target arbitrary gate types and hence, was not suitable for our study. Google's Cirq can support a variety of application unitaries and hardware gate types. However, Cirq only provides optimized decompositions for limited gate types such as SYC and CZ gates. 

\subsection{Numerical Optimization for Gate Decomposition}
Our work develops \nuop, a numerical optimization technique to efficiently decompose any application unitary into any hardware gate. \nuop~is based on recent prior work \cite{jones2018quantum,khatri2019quantum,davis2019heuristics}. We describe the method for a two-qubit application unitary and then discuss how to apply it for a program circuit. \nuop~takes an application unitary and a target hardware gate type as input and produces an optimized decomposition as output. At a high-level, \nuop~constructs a series of parameterized template circuits and optimizes each template to achieve the desired unitary. Among the optimized templates, it selects the one which satisfies an input error threshold as the final decomposition.

\begin{figure*}[t]
    \centering
    \includegraphics[width=0.8\textwidth]{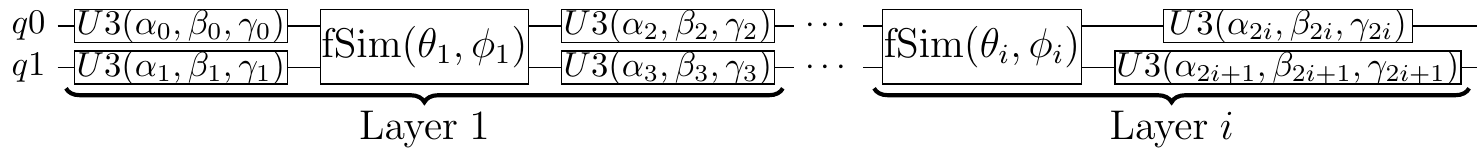}
    \caption{The template circuit used in the \nuop~ decomposition. It consists of $i$ layers, alternating parameterized single-qubit gates (three parameters $\alpha, \beta, \lambda$ for each) and fSim gates. To compile for a fixed fSim gate, $\theta$ and $\phi$ are set to the desired values in the template. To compile for Full$_\text{fSim}$, $\theta$ and $\phi$ are also considered as optimization variables.}
    \label{fig:ansatz}
\end{figure*}

Figure~\ref{fig:ansatz} shows an example template with $i$ layers. Each template circuit has alternating layers of arbitrary single qubit gates and the target hardware two-qubit gate type (similar templates are used in the ADAPT-VQE algorithm~\cite{grimsley2019adaptive}). For example, to decompose an application operation into a fSim($\pi/6, \pi/2$) hardware gate, the template has layers of fSim($\pi/6, \pi/2$) gates sandwiched between arbitary single-qubit operations. For a given template, the optimization variables are the rotation angles for the single qubit operations. By optimizing these angles and computing the accuracy w.r.t the desired unitary, \nuop\ obtains a highly accurate representation of the application operation. 
To compute accuracy, \nuop~uses \textit{decomposition fidelity} $F_{d}$, i.e., the accuracy of the unitary decomposition (without device/hardware noise), measured using the Hilbert-Schmidt inner product,
\begin{equation}
\label{eqn:hs}
F_{d}=\frac{\left \langle U_{d}, U_{t} \right \rangle_{HS}}{\mathrm{dim}(U_{t})} =\frac{\mathrm{Tr}(U_{d}^\dagger U_{t})}{\mathrm{dim}(U_{t})}.
\end{equation} 
% \resizebox{.45\vsize}{!}{$
Here, $U_t$ is the matrix of the application unitary and $U_d$ is the matrix represented by the template. Each choice of single-qubit angles in the template gives rise to a different $U_d$. When $U_d$ is far away from the desired unitary, $F_d$ is close to 0 and when it matches $U_t$, $F_d$ is close to 1, so \nuop\ maximizes $F_d$. To perform the optimization, \nuop\ uses a BFGS \cite{Flet87}, a well-known numerical optimization method.

%How does \nuop\ choose the number of template layers? 
\nuop\ starts with template circuits having one layer and grows the number of layers one at a time. For each template size $i$, it uses the optimization to obtain the best $F_d$ possible with $i$ layers. Then \nuop\ takes a fidelity threshold as input (e.g., a decomposition fidelity of 99.999\%) and selects the smallest number of layers that meets the threshold. 

\nuop\ can be flexibly used for any input gate type, as well as any application unitary by changing the input template. To evaluate Full$_{
\mathrm{XY}}$ and Full$_{\mathrm{fSim}}$ cases, i.e., the entire gate family is available, we allow \nuop~to use templates where the two-qubit gate angles are also optimization variables. Our experiments in Section \ref{sec:toolflow} confirm that the decompositions produced by \nuop\ are comparable or better than optimized KAK-based decompositions in Google Cirq. 
\begin{figure*}[t]
    \centering
    \subfloat[3-qubit circuit]{
    \includegraphics[scale=0.60]{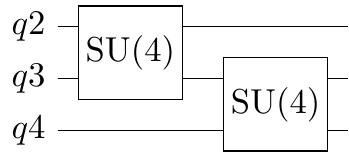}
    }
    \subfloat[Noise-aware approximate decomposition]{
    \includegraphics[scale=0.60]{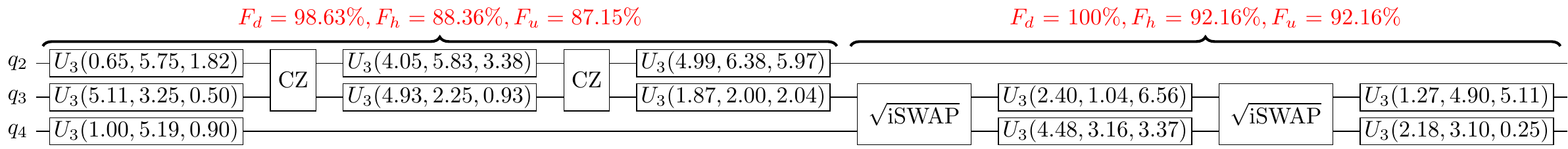}
    }
    \caption{Illustration of noise-adaptive approximate decomposition. (a) 3-qubit circuit with two gates, both having the matrix in Figure \ref{fig:qv_u1}.
    (b) decomposition circuit targeting qubits [2,3,4] on Rigetti Aspen-8 in Figure \ref{fig:rigetti_lattice} (We assume the $\sqrt{\mathrm{iSWAP}}$=XY$(\pi/2)$ gate has the same fidelity as the XY$(\pi)$ gate). 
    % The hardware gate fidelities $F_h$ of CZ and $\sqrt{\mathrm{iSWAP}}$ on qubit pair (2,3) are 0.94 and 0.70, respectively.
    The noise-aware gate compilation pass chooses the decomposition with CZ gates for qubit pair (2,3)  because its high hardware fidelity leads to higher overall fidelity $F_u=88.36\%$ (which is much higher than $F_h=0.7$ of $\sqrt{\mathrm{iSWAP}}$).
    % Only one application of CZ 
    Similarly, the decomposition with $\sqrt{\mathrm{iSWAP}}$ is selected for qubit pair (3,4).
    % where the hardware fidelity of CZ and $\sqrt{\mathrm{iSWAP}}$ fidelity are 0.93 and 0.96, respectively.
    Furthermore, the approximate decomposition maximizes the overall implementation fidelity by decomposing the unitary into only two CZ gates, compared to 3 CZ gates in the exact decomposition in Figure~\ref{fig:qv_u1_cz}.
    % (b) is a noise-unaware decomposition and (c) is a noise-aware decomposition targeting qubits [2,3,4] on Rigetti Aspen-8. XY and CZ gates are labeled with fidelity data from Figure \ref{fig:rigetti_lattice}. Both decompositions have the same number of two-qubit gates, but the noise-unaware circuit uses XY gates on qubits (2,3) which have fidelity 0.7 and has an error probability of more than 50\% in a real execution. The noise-aware decomposition uses the best gates available on each qubit pair, choosing CZ gates on qubits (2,3) and XY gates on (3,4).
    }
    \label{fig:noise_aware_example}
\end{figure*}

\subsection{Approximation and Noise-Awareness Optimizations}
Current NISQ systems have significantly higher hardware error rates (0.5-10\%) than the decomposition error thresholds commonly used for gate decomposition (e.g., $10^{-8}$ for KAK-based approaches).
%Generally, the decomposition error $\epsilon$ is extremely small (e.g. $10^{-8}$) compared to the gate error rates in NISQ systems.
%For example, IBM quantum devices have single-qubit gate error rates and two-qubit gate error rates around $10^{-3}$ and $10^{-2}$, respectively.
Therefore, recent works have observed that a less accurate decomposition that requires fewer operations can be more beneficial than an exact decomposition \cite{cross2019validating}. We incorporate this optimization into \nuop\ by considering two kinds of errors: decomposition errors and hardware errors. Decomposition error is a result of inexact representation of the target unitary. Hardware error is a result of imperfect gate implementation on a real device. For a target gate type $g_j$ and number of layers $i$, the decomposition errors are quantified using the decomposition fidelity, $F_{d}^{(g_j,i)}$ and hardware errors using hardware fidelity, $F_{h}^{(g_j,i)}$. Overall, the quality of a decomposition is the product of these two terms. By maximizing the product across gate types and the number of layers we can obtain a decomposition that has higher reliability overall, than maximizing $F_d$ alone:
\begin{equation}
\label{eqn:objective}
F_{u}^{\mathrm{APD}}=\max_{g_j, i} F_{d}^{(g_j,i)}F_{h}^{(g_j,i)}.
\end{equation}
$F_{h}^{(g_j,i)}$ is estimated by the product of the hardware fidelity (1-error rate) of every gate in the decomposition (this model has been shown to work well in real systems \cite{ncsu_iiswc, arute2019quantum, Linke3305, ionq11qubit}). 
Figure \ref{fig:noise_aware_example} shows an example decomposition with a CZ gate that has hardware fidelity $94\%$. When the above approximate optimization is used, only 2 applications of CZ are required to obtain a decomposition with overall fidelity of $88.36\%$ ($F_d = 98.63\%$). If \nuop\ had used one more CZ gate to achieves an exact decomposition (decomposition infidelity $10^{-7}$), the overall fidelity will be reduced to $83.06\%$ because of its high hardware error rate from three CZ applications ($0.94^3$).

We can further improve decompositions by exploiting fidelity variations across gate types. In Figure \ref{fig:rigetti_lattice}, on qubit pair (2,3), the CZ gate has the highest fidelity, while on qubits (3,4), the XY($\pi$) gate has the highest fidelity. When multiple gates types ($g_{j}$) are available in a given device, we can choose the best gate type for each application operation, depending on the decomposition fidelity using that gate type and it's hardware fidelity. Figure \ref{fig:noise_aware_example} shows a noise-adaptive decomposition of an application circuit using two gate types. 
%CZ and $\sqrt{\mathrm{iSWAP}}$ (we assume it has the same error rates as the XY$(\pi)$ gate). For the target SU(4) unitary, an exact decomposition requires 3 CZ or 2 $\sqrt{\mathrm{iSWAP}}$ gates. The noise-adaptive technique chooses the decomposition with CZ on qubit pair (2,3) to achieve higher overall fidelity while a noise-unaware approach will instead use the more expressive $\sqrt{\mathrm{iSWAP}}$ gate. 
We can easily implement this idea within \nuop~by computing $F_{h}^{(g_j,i)}$ using the hardware fidelity of each gate type, obtained from device calibration data. To our knowledge, our work is the first to adapt executables to noise variations across gate types.

To decompose a full application for a given instruction set having multiple two-qubit gate types, \nuop~finds the best decomposition for each application unitary and each hardware gate type, using approximation if necessary. Then, for each unitary, \nuop~selects the gate type which offers the highest overall fidelity $F_u$. \nuop~creates an output circuit where application unitaries are replaced by their best decompositions. 
%\section{Qubit Mapping with Multiple Instruction Sets} 
\section{Experimental Setup}
\label{sec:expt_setup}
{\noindent \textbf{Benchmarks:}}
We use four applications for evaluation: Quantum Volume (QV) \cite{cross2019validating}, Quantum Approximate Optimization Algorithm (QAOA)\cite{farhi2014quantum}, Fermi-Hubbard model (FH), and Quantum Fourier Transform (QFT). They cover the main types of circuits studied for QC systems i.e, random circuits (QV), quantum optimization circuits (QAOA), quantum simulation circuits (FH), and classic QC or longer-term circuits (QFT). 

QV is a full-stack benchmark proposed by IBM to compare different QC devices and compiler stacks. 
%It has been adopted by other vendors as well \cite{honeywell_qccd}. 
%QV circuits measure the size of the largest random circuit of equal width and depth that the QC system can reliably implement. 
Each random QV circuit with $n$ qubits has $n$ parallel layers of random two-qubit gates acting on a random pair of qubits.

QAOA is a promising NISQ application for solving combinatorial optimization problems. QAOA circuits have also been used to study real-system performance \cite{arute2020quantum}.
% and are similar to quantum simulation~\cite{vqe1,vqe2} circuits.
Our experiments use QAOA circuits with one layer of the MaxCut ansatz~\cite{wang2018quantum}.
% one-layer hardware efficient ansatz \cite{Moll_2018} or, having 4-6 qubits.
%  $n-1$ (hardware) and 
Each $n$-qubit QAOA circuit has $\sim n^3/4$ random two-qubit ZZ interactions, interleaved with single-qubit $X$ rotations. 

FH is a model of simulating interacting particles in a lattice, which becomes difficult to solve on classical computers for large problems. Our evaluation uses the 1D FH model circuits with one Trotter step~\cite{Arute2020ObservationOS}. Each $n$-qubit FH circuit consists of $2n$ ZZ interactions and $\sim4n$ $\frac{1}{2}$(XX+YY) interactions.

QFT is a key subroutine in conventional quantum algorithms such as quantum phase estimation and Shor's factoring~\cite{shor1994algorithms}.
An $n$-qubit QFT circuit consists of $n$ Hadamard gates and $n(n-1)/2$ $\mathrm{CZ}(\pi/2^{t})$ gates ($t\in [1, \cdots, n-1]$)~\cite{nielsen2002quantum}.

QV and QAOA circuits are generated from Qiskit~\cite{qiskit} and ReCirq~\cite{recirq}, respectively. {\em Our toolflow is scalable for large circuit sizes, but current highly noisy systems can only achieve high-fidelity computation for small-scale circuits}. Therefore, our evaluation experiments uses QV, QAOA, and QFT benchmarks with 3-6 qubits and FH circuits with 10 and 20 qubits. For each $n$-qubit QV benchmark, we use 100 random circuits with different circuit structures and unitaries to test instructions sets against a wide range of application requirements. Similarly, for each $n$-qubit QAOA benchmark we use 100 random circuits with different unitaries. 

{\noindent \textbf{Decomposition Algorithms:}} We implemented \nuop\ as a IBM Qiskit compiler pass using version 0.20. Numerical optimization was performed using the BFGS implementation \cite{Flet87} in scipy version 1.4.1. In \nuop\ we used templates with a maximum of ten two-qubit gate layers, but for most cases less than four layers were sufficient for obtaining high quality decompositions. To validate \nuop\, we compare it to the decomposition passes in Google Cirq v0.8.2.

{\noindent \textbf{QC Systems and Simulations:}} We use realistic noise simulations of two systems: Rigetti Aspen-8 and Google Sycamore. Aspen-8 has 30 qubits and has 4 connected rings with 8 qubits each (two qubits are not functional). Figure \ref{fig:rigetti_lattice} shows the first ring of qubits. Sycamore is a 54-qubit device having grid connectivity. 
For Aspen-8 we obtain real calibration data from~\cite{rigetti_qcs}, including error rates for each gate type (XY$(\pi)$, CZ, single-qubit), coherence times and readout errors. 
%We use the data from July 30, 2020.  
For experiments with arbitrary XY($\theta$) gates, we model the gate errors rates as a uniform distribution in the range 95-99\% (based on experiments in \cite{abrams2019implementation}).
For Sycamore we obtain qubit coherence times and readout errors from \cite{arute2019quantum} and acquire error rates of simultaneous SYC gates and single-qubit gates via Google's quantum computing service. We model the error rates of other types of two-qubit gates as a normal distribution with $\mu=0.62\%, \sigma=0.24\%$ (based on the error rate distribution of SYC gates). For both systems, we use the Qiskit Aer simulator \cite{qiskit} to perform a noise simulation using the calibration data. 
Specifically, it applies single-qubit and two-qubit depolarizing noises based on single-qubit and two-qubit gate error rates. It implements amplitude damping and dephasing noise based on T1 and T2 times as well as gate duration.

{\noindent \textbf{Metrics:}} For QV, the standard metric is the average \textbf{probability of heavy output generation (HOP)} obtained using 100 random circuits \cite{cross2019validating, aaronson2016complexity}. With $n$ qubits, HOP greater than 2/3 denotes that the set of qubits has quantum volume $2^n$. For QAOA benchmarks, we measure the average \textbf{cross-entropy difference (XED)}~\cite{Boixo2018} metric which tests the quality of the generated output with respect to the ideal distribution (noiseless simulation) and the uniform distribution.
% Higher values of cross-entropy difference are better. 
For FH benchmarks, we use the linear cross-entropy benchmarking \textbf{fidelity} \cite{Neill195} metric to compare the measured probabilities with the ideal probabilities.
For QFT benchmarks, we measure the \textbf{success rate}, i.e., the probability of an execution getting correct quantum states. Higher values are better for all three metrics. For all circuits, we perform simulations for 10000 shots. We also report the \textbf{two-qubit instruction counts} required to express an application using a particular gate set.
% For QAOA circuits we also use \textbf{L1 Norm distance (LND)}, which is a stricter metric than XED and smaller values are better. 

{\noindent \textbf{Compilation setup and performance:}}
% \textcolor{blue}{NuOp’s compilation time scales as $O(GH)$ where $G$ is the number of 2-qubit gates in the application and $H$ is the number of available hardware 2-qubit gates. The compile time is independent of the number of qubits.} 
Our experiments were performed on two systems: Intel Xeon Gold 6230 CPU 2.10GHz (using up to 32 threads, 128GB RAM) and Intel Xeon Gold 6140 CPU 2.30GHz (using up to 36 threads, 192GB RAM).
% We use a parallel implementation of \nuop~(32 threads) and all benchmark circuits in our study can be compiled in under tens of seconds.
NuOp’s compilation time scales as $O(GH)$ where $G$ is the number of 2-qubit gates in the application and $H$ is the number of available hardware 2-qubit gates. The compile time is independent of the number of qubits.
Since decomposing an application gate to a hardware gate is independent of other gates, we use a parallel implementation of NuOp. With 32 threads, decomposing a circuit with 1000 2-qubit gates to one hardware gate type requires around 220 seconds.

\section{Toolflow Validation and Tuning}
\label{sec:toolflow}
\begin{figure}[t]
    \centering
    \includegraphics[scale=0.45]{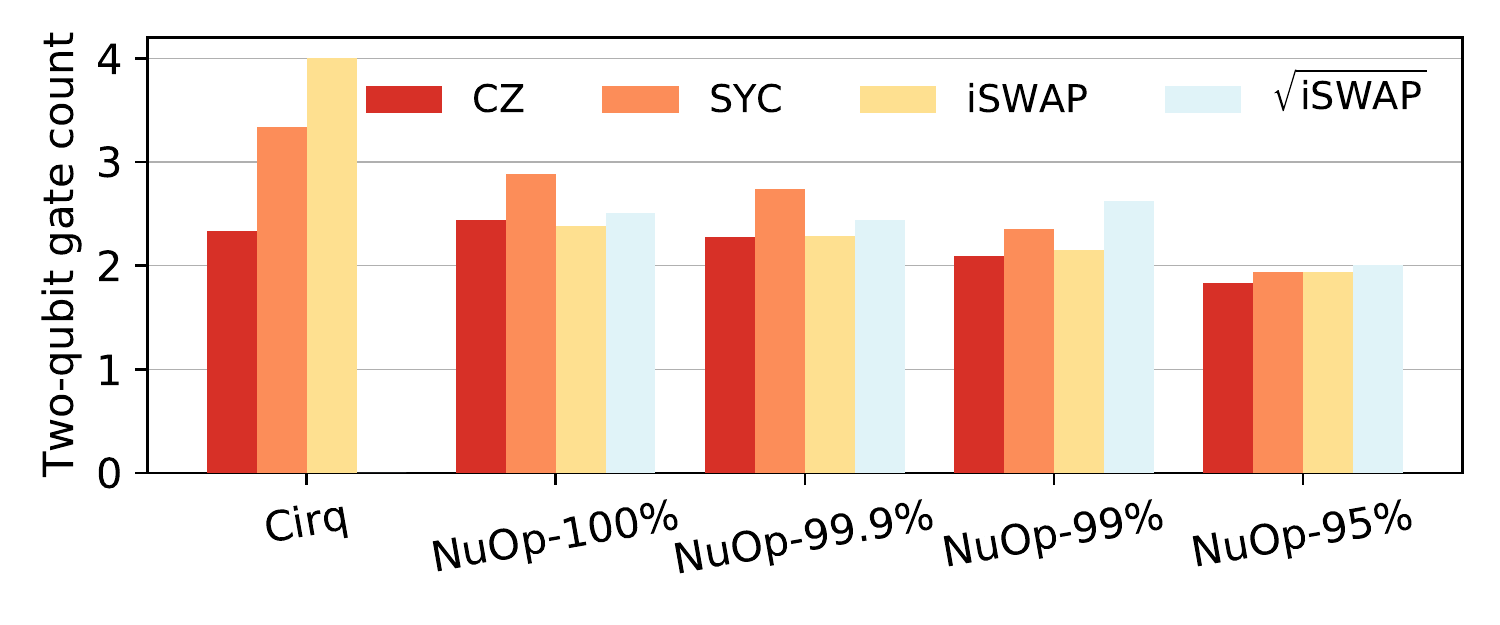}
    \caption{Comparison with Cirq. Hardware gate counts (lower is better) averaged across 
    %100 random 
    applications. Legend shows the target two-qubit gate types. \cirq~does not support decompositions for QV with $\sqrt{\mathrm{iSWAP}}$. \nuop-99.9\% denotes that $99.9\%$ is the target hardware fidelity for the decomposition. On average, \nuop\ uses 1.3X-2.3X less gates than \cirq, depending on the hardware gate fidelity and gate type. For \nuop, the average decomposition errors are within a specified and tunable threshold based on the device requirements, e.g., for \nuop-99\%, the average decomposition error is 0.12\%.
    }
    \label{fig:expt_gate_choices}
\end{figure}
\begin{figure}[t]
    \centering
    \includegraphics[height=0.21\textwidth]{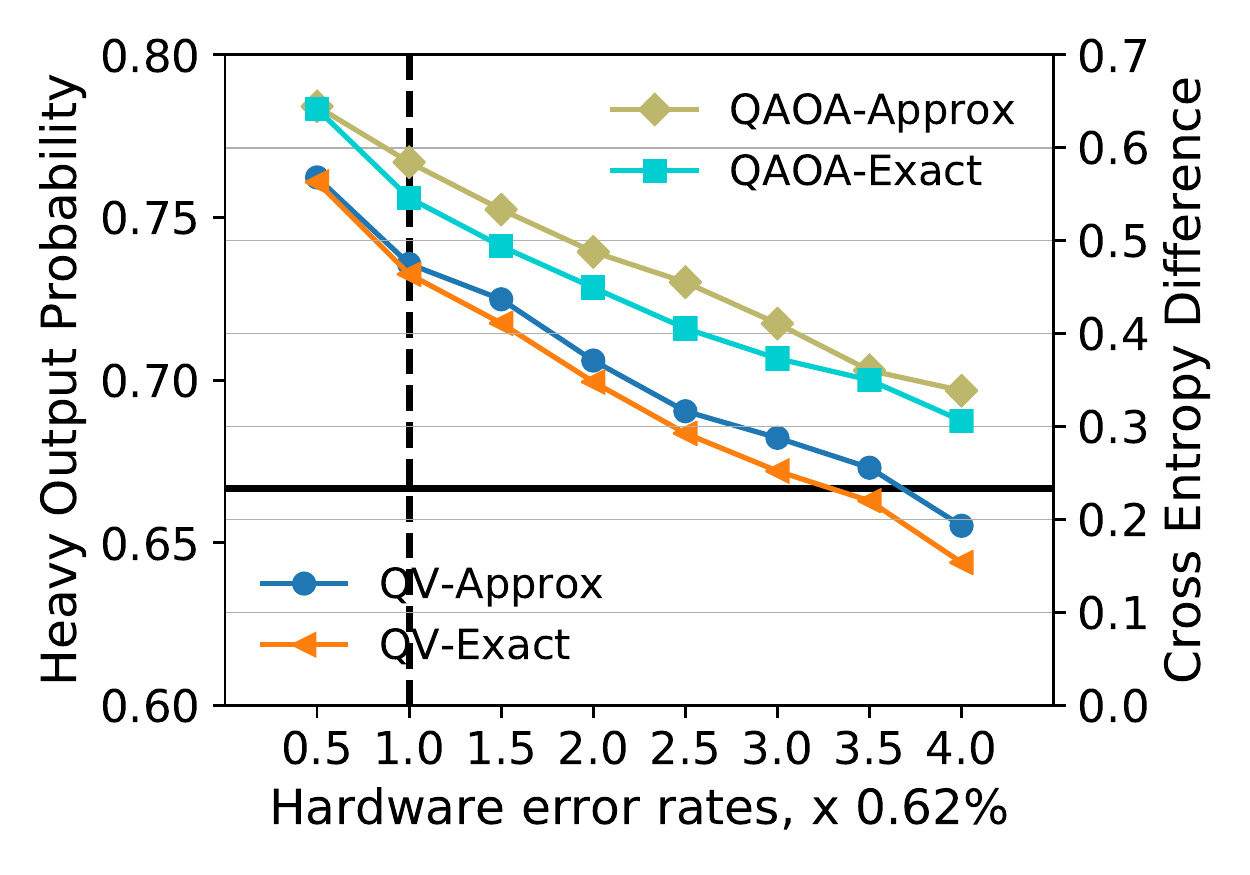}
    \caption{Comparison of exact and approximate decomposition on application performance with different average error rates of the SYC gate. 
    The left and right y axis denote the average HOP of 100 5-qubit QV benchmarks and the average XED of 100 4-qubit QAOA benchmarks respectively (higher values are better).
    In the low noise regime, applications that are decomposed by the approximate approach have similar performance as the ones using exact decomposition.
    Approximate decomposition starts outperforming the exact approach when the device has average error rate around Sycamore's (dashed vertical line).}
    \label{fig:exact_approximate}
\end{figure}

\subsection{Comparison of \cirq~and exact decomposition with \nuop}
% {\noindent \textbf{Comparison of \cirq~and \nuop:}}
Figure \ref{fig:expt_gate_choices} compares the average hardware gate counts and decomposition error rates for 100 random QV, QAOA, and QFT unitaries using the gate types from Table \ref{fig:systems_gate_sets}. 
Compared to the KAK-based decomposition in \cirq, exact decomposition with \nuop~i.e., \nuop-100\% finds decompositions with similar or less amount of applications of two-qubit hardware gates. For example, \nuop~uses 3 CZ, 3 SYC, or 3 iSWAP gates for one QV unitary while \cirq~requires 3 CZ, 6 SYC, or 4 iSWAP gates. Across applications, \nuop-100\% reduces gate counts by 1.26X on average, compared to Cirq.
Furthermore, \nuop~can decompose any target unitary with any native gate while KAK-based decomposition like \cirq~is typically implemented for decomposing unitaries into specific native gates. For example, \cirq~does not support decompositions of QV unitaries with  $\sqrt{\mathrm{iSWAP}}$. Due to \nuop's flexibility in hardware gate sets and gate reductions over \cirq, we use it for all other experiments in this work.

\subsection{Comparison of exact and approximate decomposition}
From Figure~\ref{fig:expt_gate_choices}, across gate sets, \nuop~offers 1.33-1.68X average reduction in gate counts compared to Cirq when approximation is applied. Compared to \nuop-100\%, the reductions with approximation versions of \nuop~ are average 1.05-1.33X.
%For QV unitaries, \nuop-99.9\% (\nuop-99\%) produces decompositions that have 2.7 (2.3?) CZ gates on average (10\% (23\%) reduction compared to \nuop-100\%), with around $0.01\%$ ($0.1\%$) decomposition infidelity (i.e., inexact representation of the unitary, not hardware error). With higher hardware error thresholds, approximate decomposition offers more gate reduction. 
For example, \nuop-95\% obtains QV decompositions that have an average of 1.8 CZ gates per unitary (40\% lower than \nuop-100\%) with decomposition infidelity around 3\% (not shown). %These reductions in instruction count lead to improvements in fidelity. 
%In NISQ systems that have high two-qubit gate errors, this reduction in gate count can lead to improvements on application performance.
Figure~\ref{fig:exact_approximate} shows that for both QV and QAOA benchmarks, decomposition with the approximate approach achieves similar application performance as exact decomposition in less noisy systems and outperforms exact decomposition for systems with higher error rates (around 0.62\%).
Given that the average two-qubit gate error rates in the Sycamore processor and the Aspen-8 processor are 0.62\% and 95\%, we use the approximate decomposition approach for evaluating different instruction sets on these two devices.
% For systems like Aspen-8, such operating points are useful because the hardware error rates of two-qubit gates are high and variable across qubits. 

\begin{figure*}[t]
    %\centering
    \raggedright
    \subfloat[QV unitaries]{
    \includegraphics[scale=0.27]{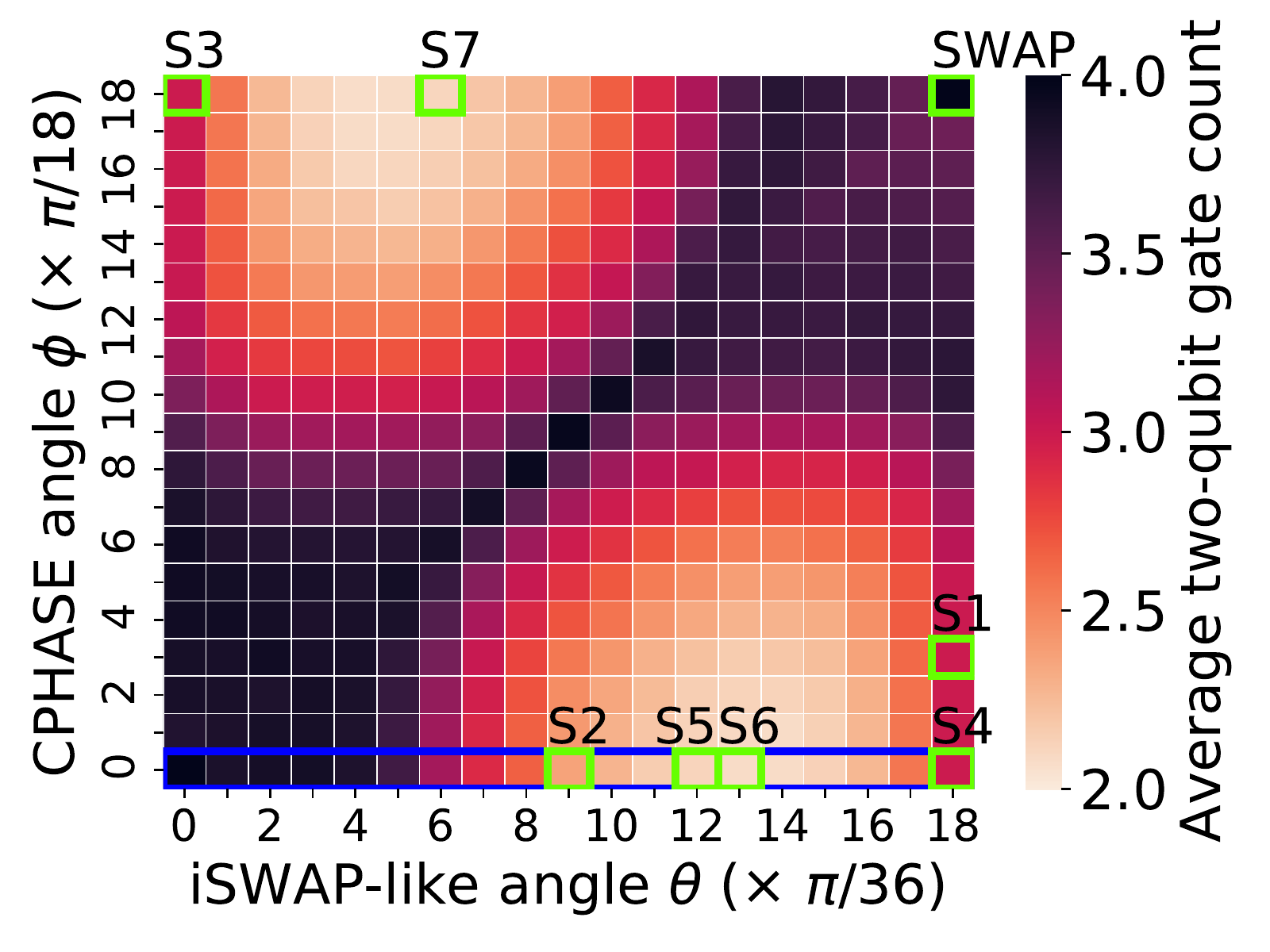}
\label{fig:qv_heatmap}    
    }
    \subfloat[QAOA unitaries]{
    \includegraphics[scale=0.27]{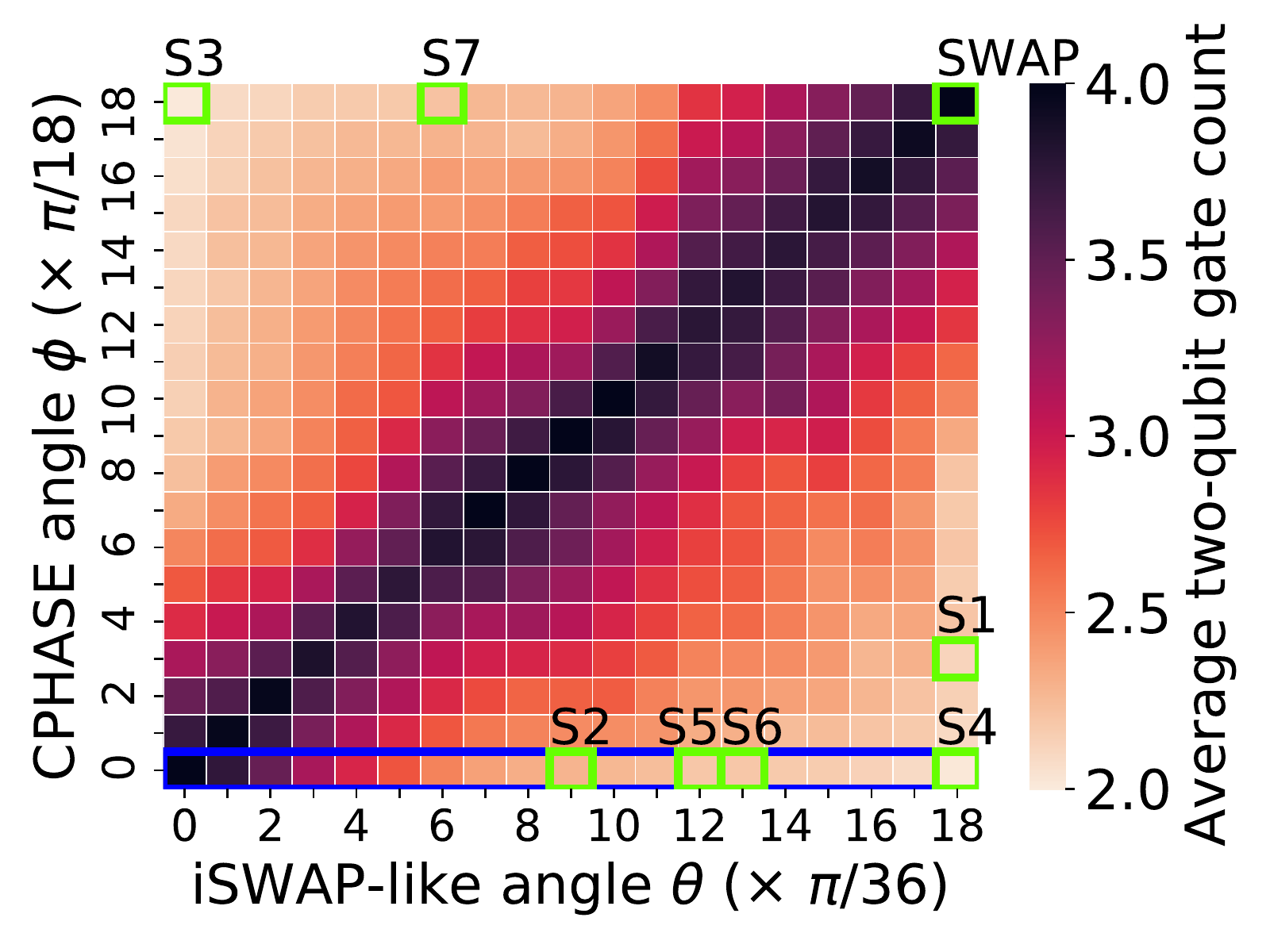}
\label{fig:qaoa_heatmap}    
    }
    \subfloat[QFT unitaries]{
    \includegraphics[scale=0.27]{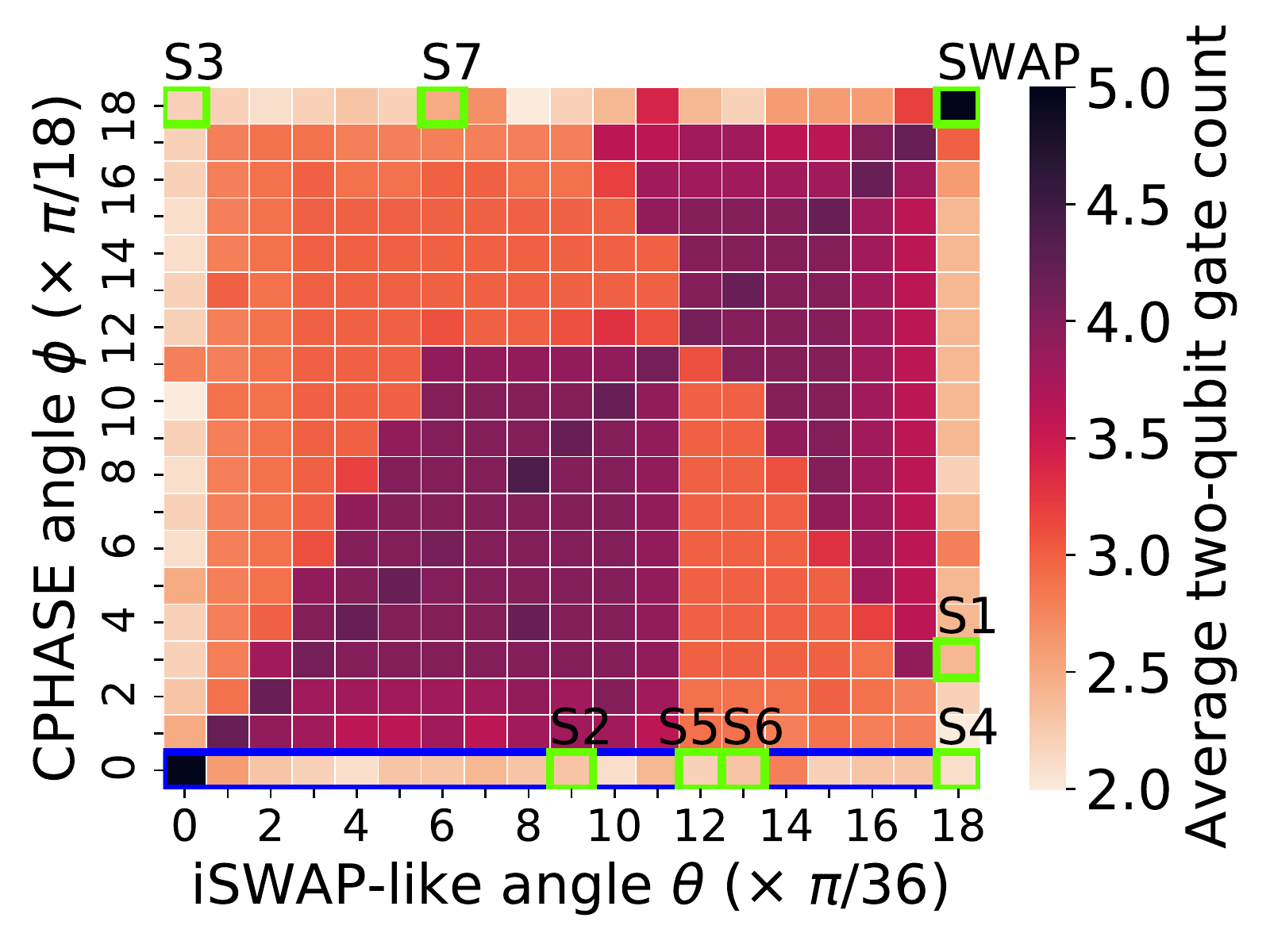}
\label{fig:qft_heatmap}    
}
    \subfloat[FH unitaries]{
    \includegraphics[scale=0.27]{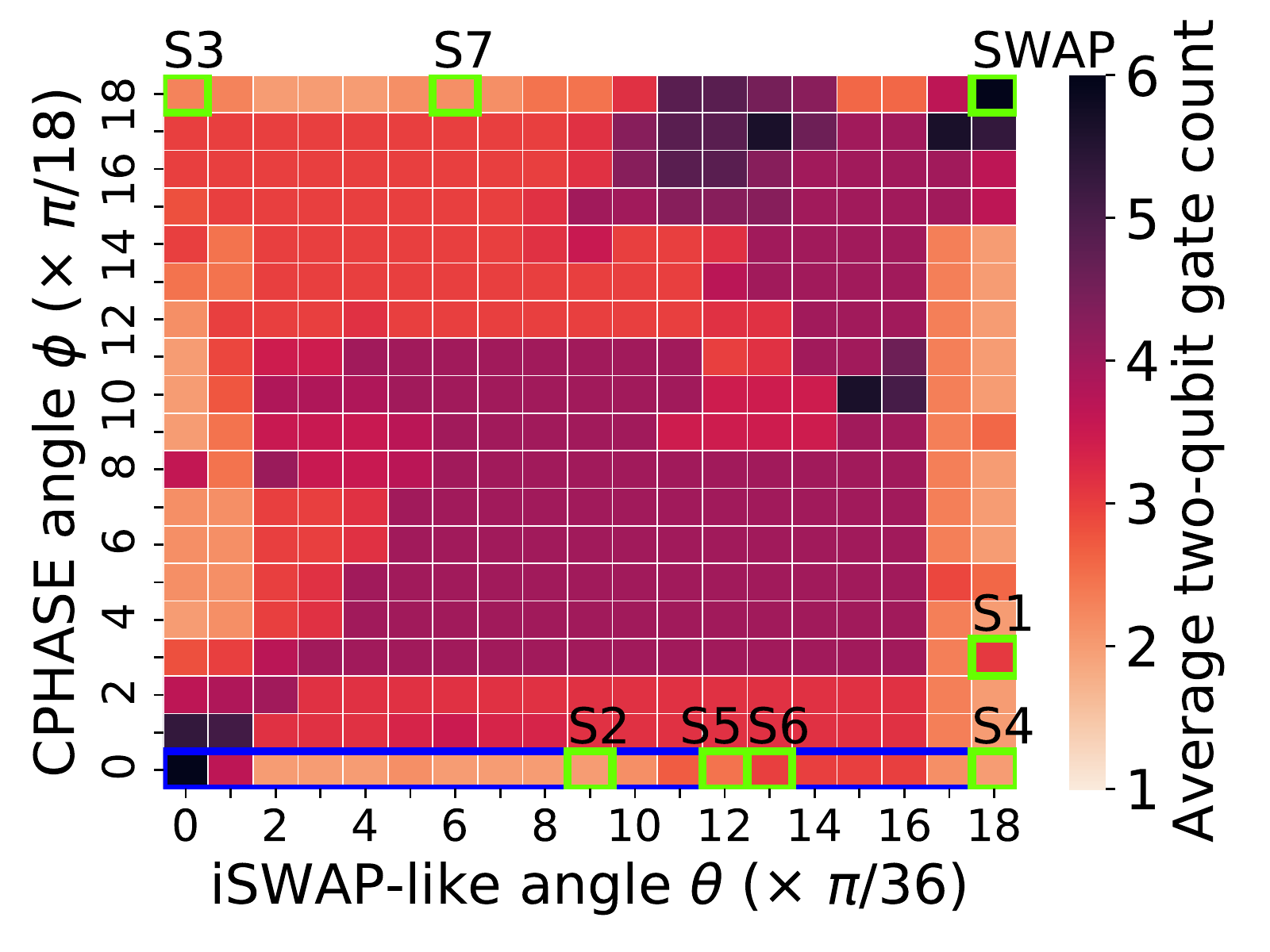}
\label{fig:fh_heatmap}    
}

    \subfloat[SWAP unitary]{
    \includegraphics[scale=0.27]{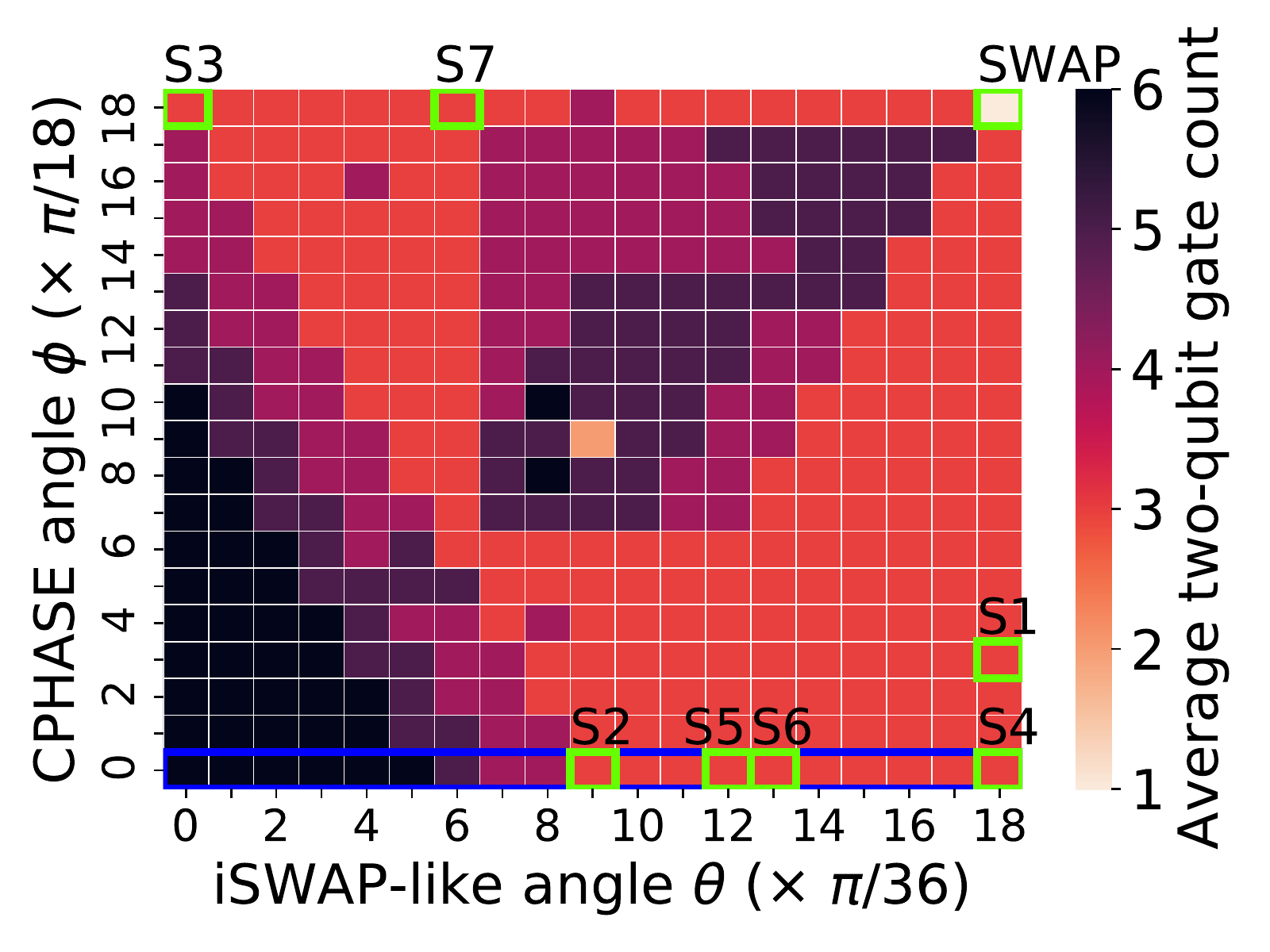}
\label{fig:swap_heatmap}    
    }    

  % The caption
  \vspace*{\dimexpr-\parskip-110pt\relax}% Skip backwards over last left-aligned image
  \parshape 8 % Set flow of caption: 6 lines...
    .27\textwidth .73\textwidth % First 5 start @ .5\textwidth with
    .27\textwidth .73\textwidth % a width of .5\textwidth
    .27\textwidth .73\textwidth
    .27\textwidth .73\textwidth
    .27\textwidth .73\textwidth
    .27\textwidth .73\textwidth
    .27\textwidth .73\textwidth    
    .27\textwidth .73\textwidth    
    %0pt \textwidth % last (sixth) line restores regular flow ad infinitum
  \makeatletter
  % Setting of actual caption (this is taken from latex.ltx)
  \refstepcounter\@captype% Increase float/caption counter
  \addcontentsline{\csname ext@\@captype\endcsname}{\@captype}% Add content to "List of..."
    {\protect\numberline{\csname the\@captype\endcsname}{ToC entry}}%
  \csname fnum@\@captype\endcsname: % Float caption + #
  \makeatother
Average two-qubit hardware gate counts (lower is better) when decomposing (a) 1000 random QV unitaries, (b) 1000 random QAOA unitaries, (c) 10 QFT unitaries, (d) 60 FH unitaries, and (e) the SWAP unitary into different gates in the $\mathrm{fSim}(\theta,\phi)$ family using \nuop's exact decomposition ($F_d \leq 10^{-6}$). The XY gate set is a subset of the fSim gate set and is highlighted by the blue box. We select the types S1-S7 and the hardware SWAP gate, highlighted by green boxes. These types are used as baseline instruction types in Table II, and their combinations are used to study instruction sets with mutiple gate types. If all gates in the fSim set are available (i.e. a continuous set), then the average gate counts for QV, QAOA, QFT, FH and SWAP unitaries are 2, 1, 1, 1 and 1, respectively. 
    \label{fig:fsim_heatmap}
\end{figure*}

\begin{figure*}[t]
    \centering
    \subfloat[100 3-qubit QV circuits (higher is better)]{
        \includegraphics[scale=0.5]{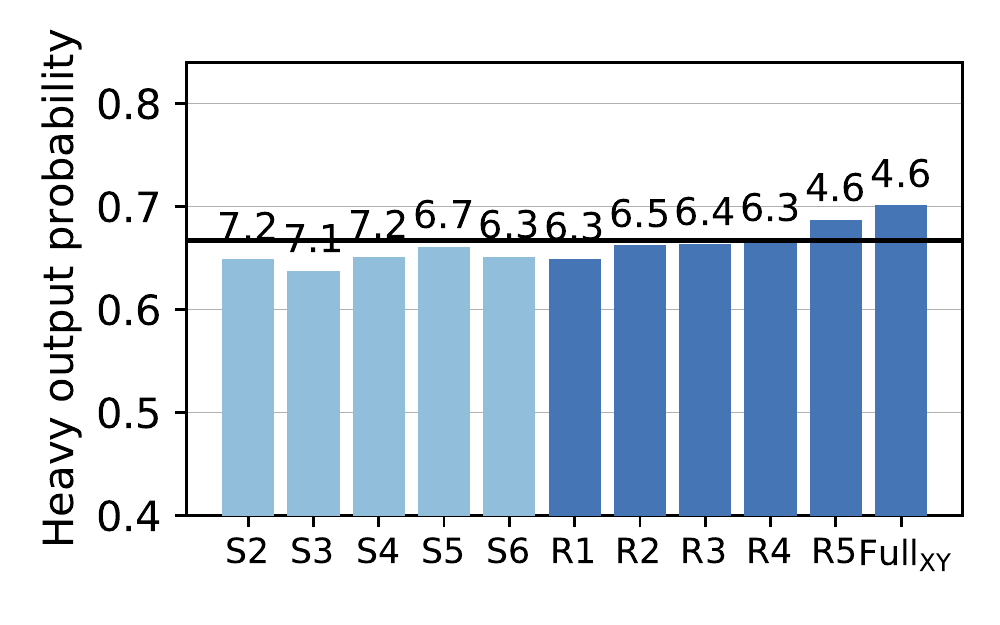}
        \label{fig:rigetti_qv}    
    }
    \subfloat[100 4-qubit QAOA circuits (higher is better)]{
        \includegraphics[scale=0.51]{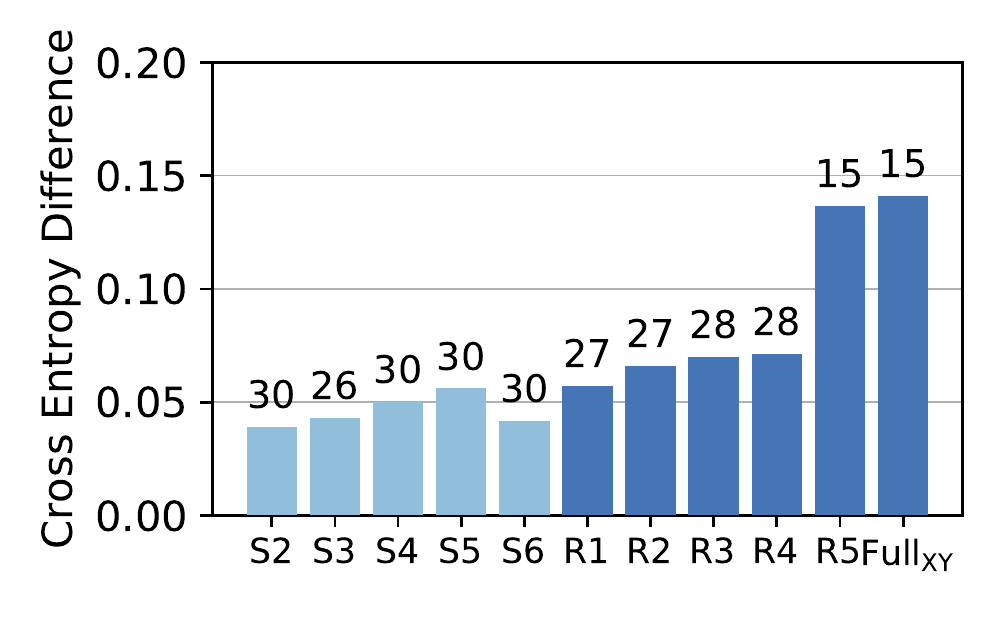}
        \label{fig:rigetti_qaoa}
    }
    \subfloat[The 3-qubit QFT circuit (higher is better)]{
        \includegraphics[scale=0.5]{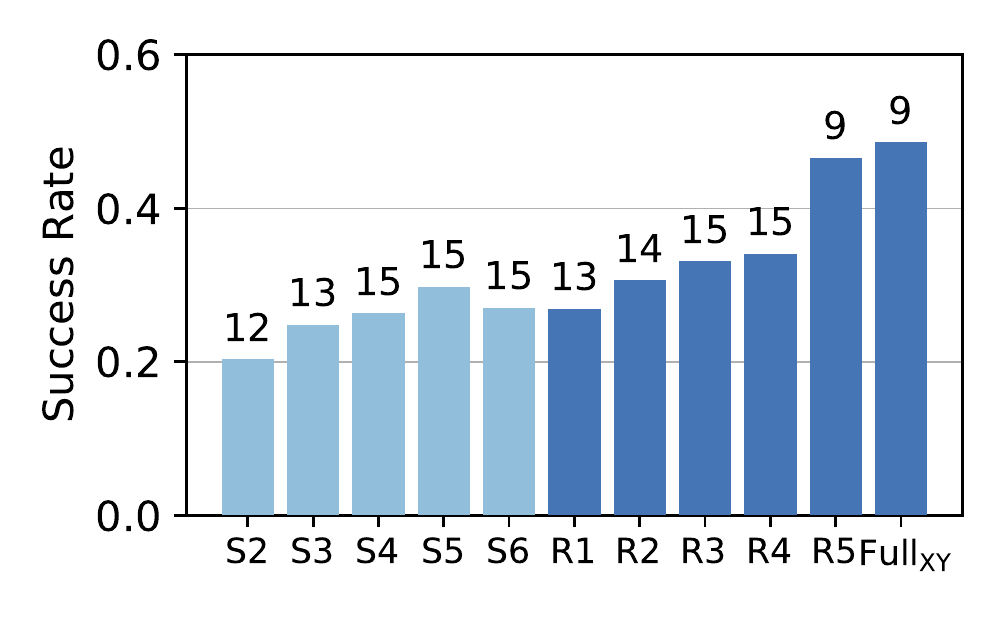}
        \label{fig:rigetti_qft}
    }
    \caption{Simulation results on Rigetti Aspen-8 with noise variations across gate types. The QV threshold is marked with a horizontal line at \textcolor{black}{2/3}. Instruction sets with multiple gate types (R1-R5) improve application performance compared to single-gate instruction sets (S2-S6). R5 that includes hardware SWAP gate, nearly achieves the same maximum reliability as  Full$_{\mathrm{XY}}$. }
    \label{fig:rigetti_sim}
\end{figure*}

\begin{figure*}[t]
    \centering
    \subfloat[100 6-qubit QV circuits (higher is better)]{
        \includegraphics[scale=0.45]{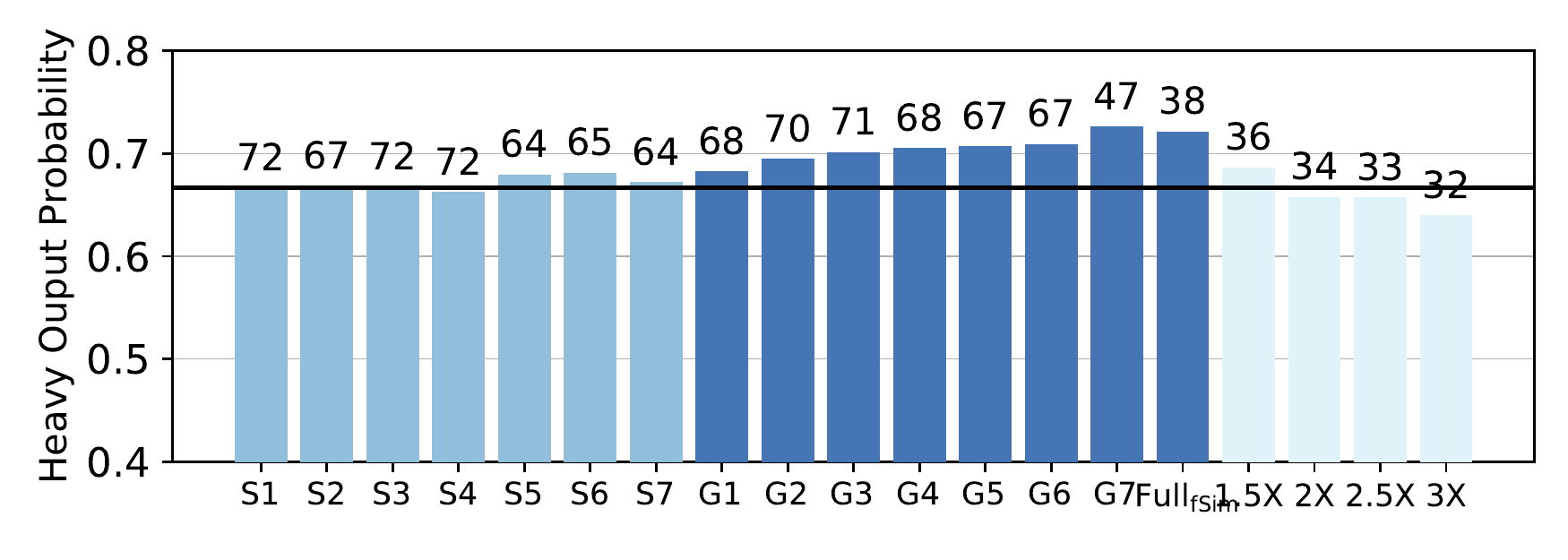}
        \label{fig:sycamore_qv_na}    
    }
    \subfloat[100 6-qubit QAOA circuits (higher is better)]{
        \includegraphics[scale=0.45]{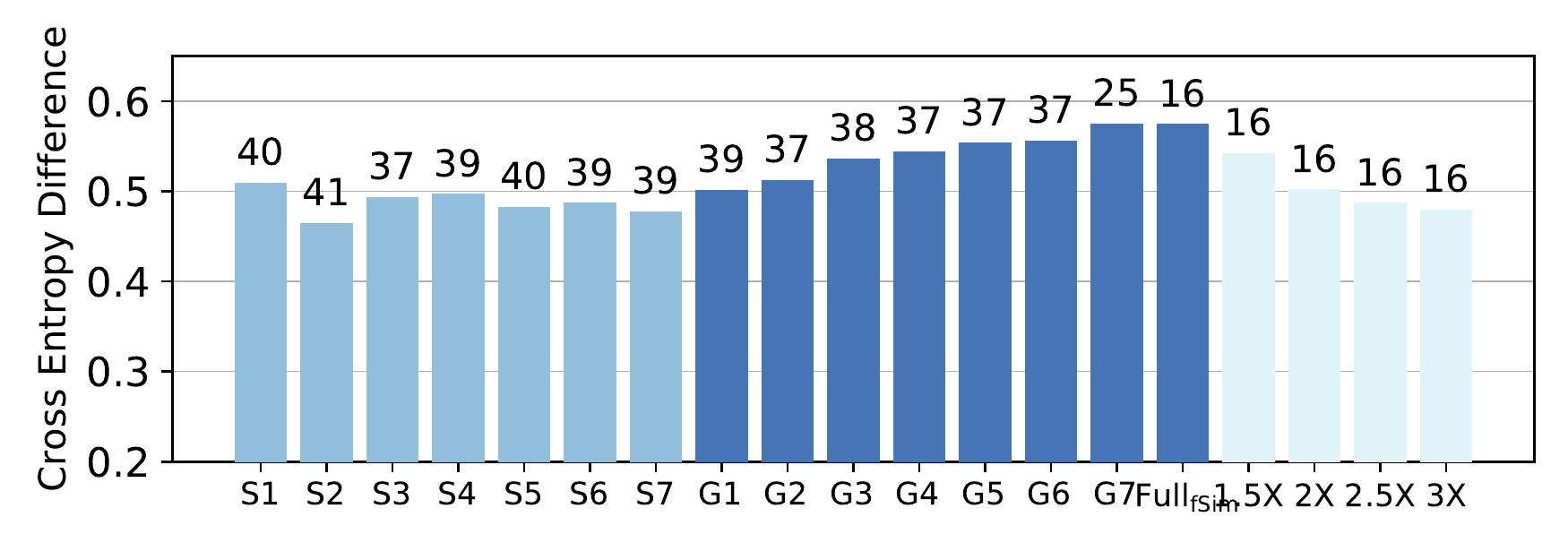}
        \label{fig:sycamore_qaoa_na}
    }

    \subfloat[The 6-qubit QFT circuit (higher is better)]{
        \includegraphics[scale=0.45]{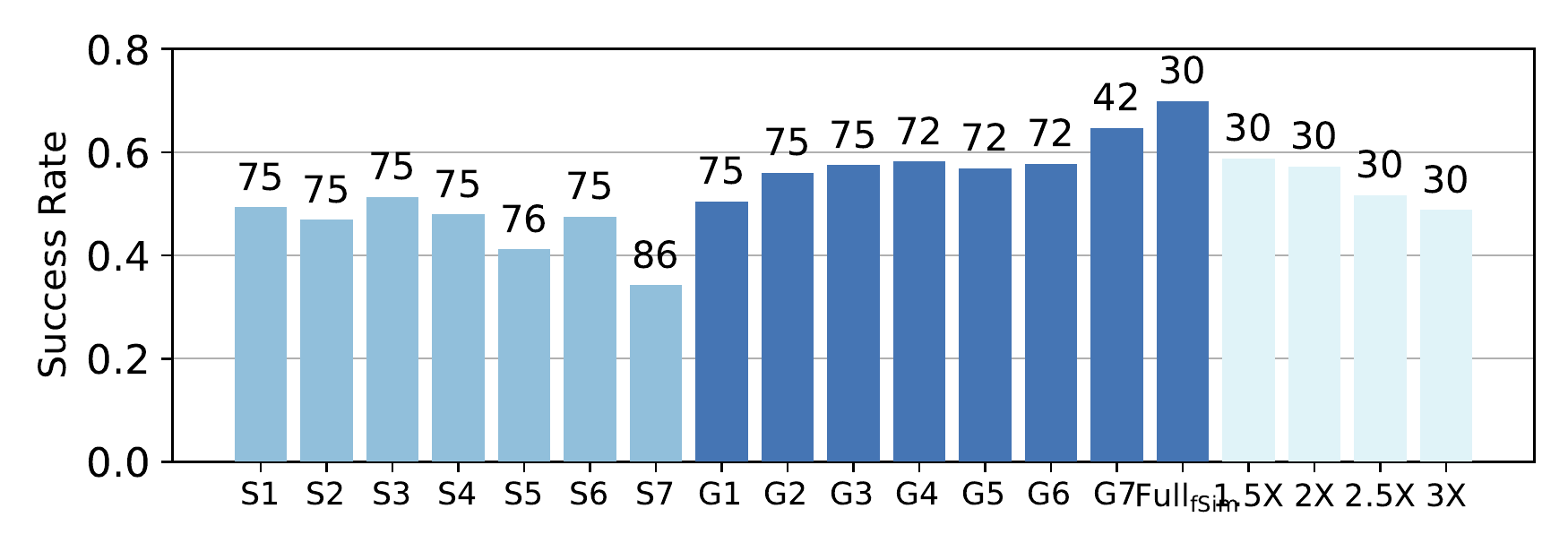}
        \label{fig:sycamore_qft_na}
    }
    \subfloat[The 10-qubit FH circuit (higher is better)]{
        \includegraphics[scale=0.45]{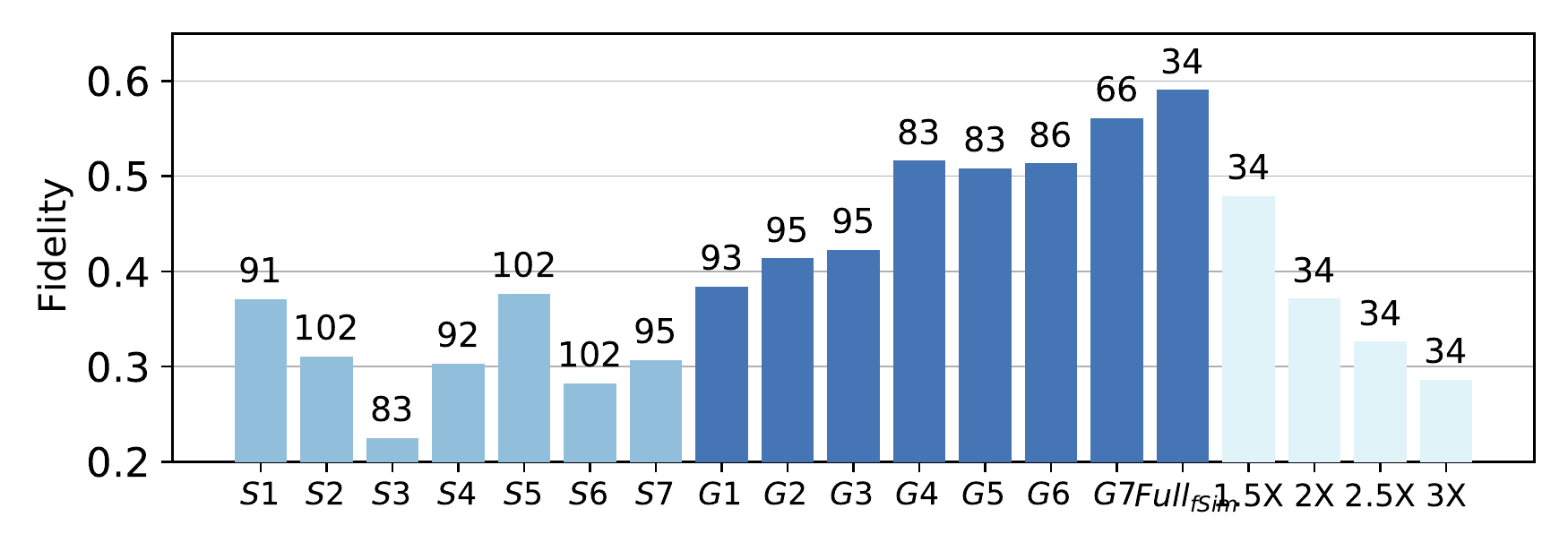}
        \label{fig:sycamore_fh_na}
    }
    
    \subfloat[100 6-qubit QAOA, no noise variations across gate types]{
    \includegraphics[scale=0.5]{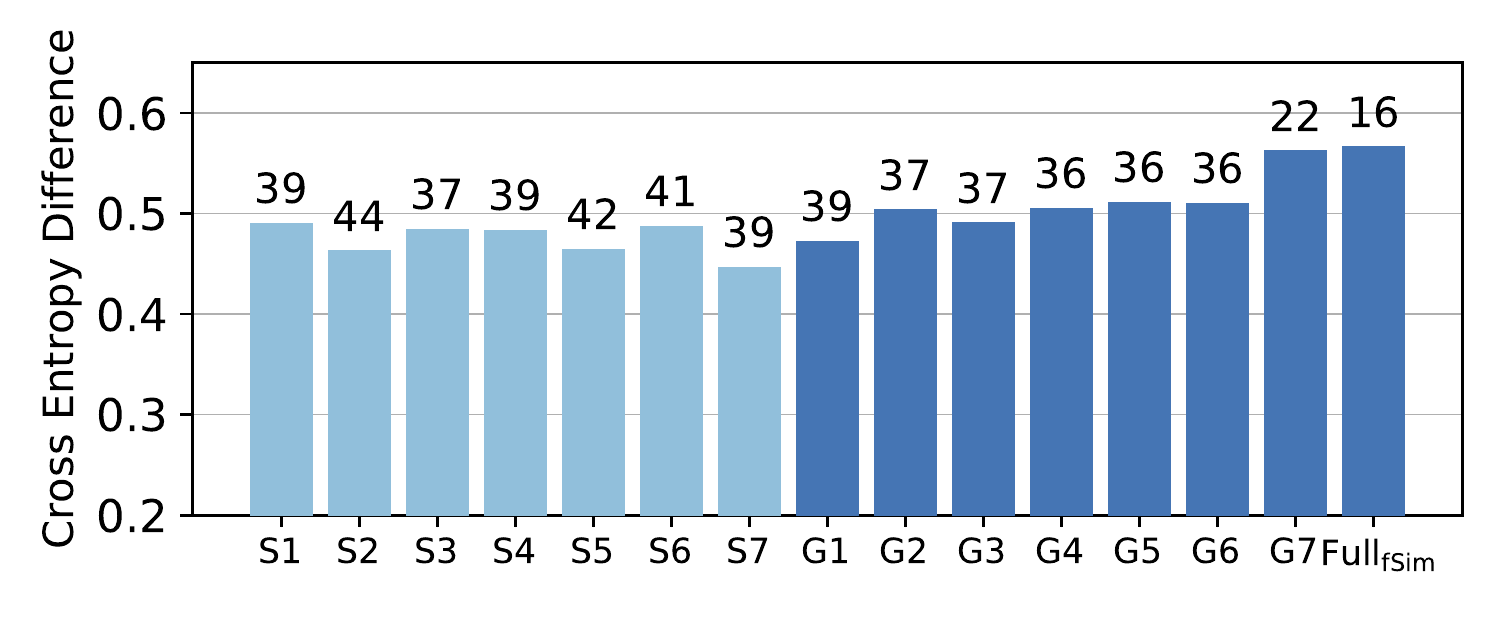}
    \label{fig:sycamore_qaoa_nu}
    }
    \subfloat[10- and 20-qubit FH with different noise levels]{
    \includegraphics[scale=0.5]{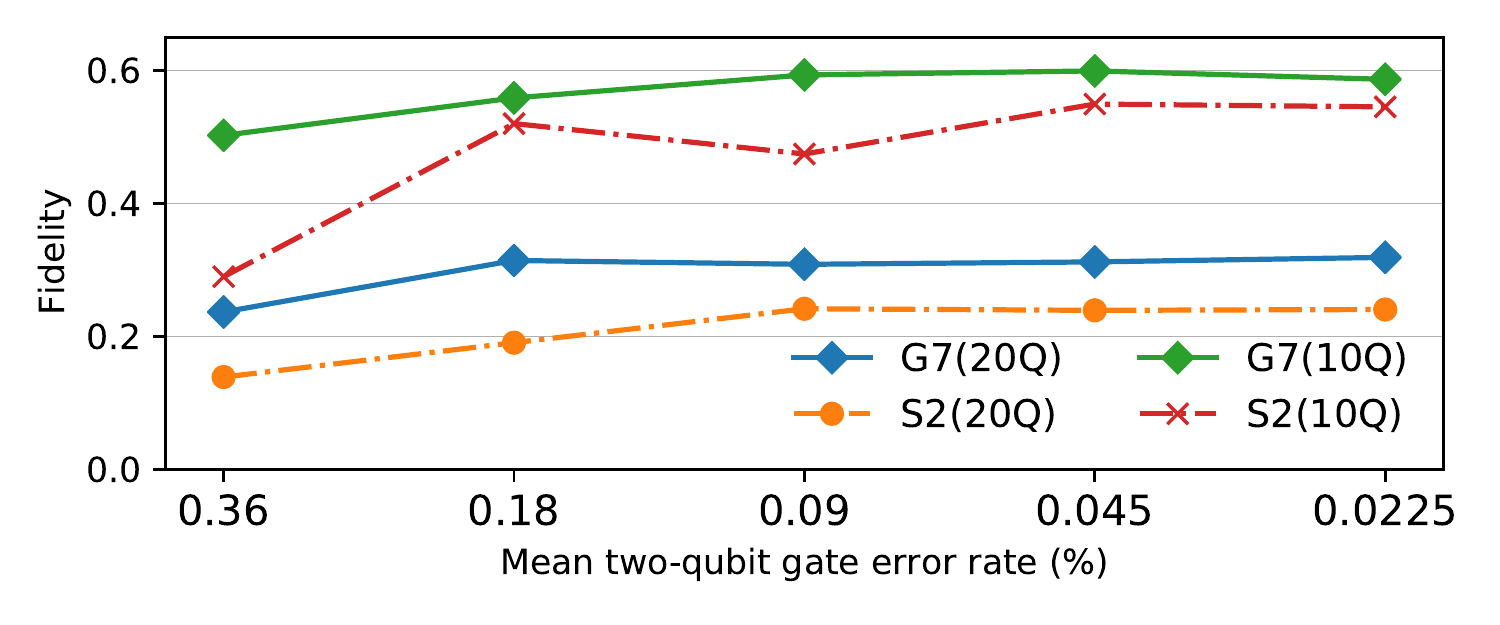}
    \label{fig:fh}
    }
    \caption{Simulation results on Google Sycamore (a-d) with and (e) without noise variation across gate types. The QV threshold is marked with a horizontal line at \textcolor{black}{2/3}. Multi-gate instruction sets (G1-G7) improve application performance compared to single-gate instruction sets (S1-S7). Compared to (b), the improvements of G1-G7 in (e) are less if no noise variation across gate types. The instruction set with SWAP gate (G7) achieves comparable reliability as  Full$_{\mathrm{fSim}}$. The advantages of the continuous set reduces or even disappear when the average gate error rates in continuous set increase to 2X (the lightest blue bars). (f) Fidelity for Fermi-Hubbard model circuits with 10 and 20 qubits. Average two-qubit hardware error rates are varied from  0.02\% (low noise) to 0.36\% (high noise, current error rate). Across circuit sizes and noise ranges, multiple gate types (G7) outperforms the single gate type.}
    \label{fig:sycamore_na}
\end{figure*}

\section{Benefits of Multiple Gate Types}
\label{sec:benefits}
\subsection{Optimizing Instruction Count}
% {\noindent \textbf{Expressivity of different hardware gates:}}
Figure \ref{fig:fsim_heatmap} shows the expressivity for application unitaries across the parameter space of XY and fSim gates. We uniformly discretized the continuous space into a $19\cross19$ grid of $\theta$ and $\phi$ values for this experiment. The value range of $\theta$ and $\phi$ are set to be $[0,\pi/2]$ and $[0,\pi]$ (instead of $[0,2\pi]$) because of the unitary symmetries under single-qubit phase rotations. For each parameter combination, we calculated the average two-qubit instruction counts for 1000 random QV unitaries, 1000 random QAOA unitaries, 10 QFT unitaries, 60 FH unitaries, and the SWAP unitary.
% For each parameter combination, the total two-qubit instruction count was computed across a range of two-qubit unitaries for each application. 
The heatmap shows these instruction counts normalized by the number of instructions in the application i.e., the number of two-qubit instructions per application operation. 
Overall, the average instruction counts are strongly influenced by the application and gate parameter choices. For each application, instruction counts vary between 1 to 6 depending on the gate parameters. Across gate types and applications, these large differences in instruction counts imply that choosing the right hardware gate types, based on the application requirement is important for NISQ systems.

For QV unitaries, gates that are close to $\mathrm{fSim}(5\pi/12,0)$ and $\mathrm{fSim}(\pi/6,\pi)$ are most expressive and near to the optimal value with a continuous set, i.e., only around 2 instructions are required per unitary (compared to 3 applications of CZ, SYC, or iSWAP). In contrast, the most expressive gates for QAOA unitaries are around iSWAP and CZ (2 applications per unitary). 
For QFT unitaries, 
% most iSWAP-like gates ($\phi=0$), CPHASE gates ($\theta=0$),
most $\mathrm{fSim}(\theta,\pi(0))$ and $\mathrm{fSim}(\pi/2(0),\phi)$ gates are expressivity-efficient (2 applications per unitary).
For FH unitaries, gates close to $\sqrt{\mathrm{iSWAP}}$ are more expressive (2 applications per unitary). We also studied the instruction counts required for the SWAP unitary, since SWAPs are the fundamental means of qubit movement in superconducting systems. 
For a large fraction of fSim gates, 3 gate applications are required to construct one SWAP unitary. With fsim$(\pi/4, \pi/2)$, we require 2 gate applications and with fSim($\pi/2, \pi$), we require only 1 gate application since this gate is equivalent to the SWAP gate up to single-qubit rotations.

From these heatmaps, %we also observe that
certain gate types perform well for multiple applications. For example, fSim($\pi/6, \pi$) gives low instruction counts for all cases. Selecting gates in this manner, we arrived at a set of seven two-qubit gate types that offer low instruction counts. These types are shown in Table \ref{tab:instruction}. This list also includes CZ, XY($\pi$), SYC, and $\sqrt{\mathrm{iSWAP}}$ gates which are used in current Rigetti and Google systems. We use these gate types and their combinations for the subsequent experiments.

\subsection{Optimizing Application Reliability}

Figures \ref{fig:rigetti_sim} and \ref{fig:sycamore_na} show the application reliability comparison among instruction sets with a single gate type (S1-S7), instruction sets having multiple gate types (R1-R5 for Rigetti and G1-G7 for Google), and instruction sets which use the full continuous parameter space (Full$_{\mathrm{XY}}$ for Rigetti and Full$_{\mathrm{fSim}}$ for Google). In Figure \ref{fig:rigetti_qv} and \ref{fig:sycamore_qv_na}, instruction sets with multiple types increase the heavy output probability (HOP) by up to 6.6\% (on average 2.3\%), compared to the best instructions sets with one type. HOP is averaged across a large number of circuits and even small improvements are very significant for applications \cite{cross2019validating}. In particular, for Rigetti, none of the single gate type instruction sets reach the HOP threshold of $2/3$, while R3, R4 and R5 cross the threshold. That is, having access to these 4-6 types of two-qubit gates can double the quantum volume of the system. When the instruction set has multiple gate types, the cross-entropy differences of QAOA benchmarks improve by up to 12.9\% (5.9\% on average) in Google and 143.7\% (43.1\% on average) in Rigetti. Similarly for QFT, the success rates increase by up to 26\% (11.6\% on average) in Google and 56.5\% (15.2\% on average) in Rigetti. Finally, the fidelity of the FH circuit increases by up to 56.8\% (29.8\% on average) in Google. 
We further compare the application reliability using multiple-gate sets with the values achieved by using full continuous sets.
% where the application has access to any fSim gate parameter combination.
% We compared the fidelity of G1-G7 to fSimFull where the application has access to any fSim gate parameter combination.
In Figure \ref{fig:rigetti_sim} and \ref{fig:sycamore_na}, across benchmarks and quantum systems, the applications decomposed with Full$_{\mathrm{XY}}$ and Full$_{\mathrm{fSim}}$ have much higher reliability than the decomposition with R1-R4 and G1-G6 instruction sets. 
This is because many routing operations (i.e., application SWAP gates in Qiskit compiler) are inserted to comply with the connectivity limitation in Sycamore and Aspen-8. This leads to a significant increase in the total instruction counts. 
%However, compared to Full-fSim, gates in sets R1-R4 and G1-G6 are only comparably expressive for QV, QAOA, and QFT unitaries but much less efficient for the SWAP unitary.
These SWAP operations are difficult to express succintly using the gates in R1-R4 and G1-G6. When the SWAP gate is added as a native type, i.e., in R5 and G7, the instruction set offers reliability close to the optimal values achieved by Full$_{\mathrm{fSim}}$. \emph{Therefore, our work shows that implementing multiple types of two-qubit gates is beneficial and only a few application-expressive types are required to obtain reliability comparable to using the entire continuous set.}
% Similarly, for Rigetti, Rx and Ry obtain near-optimal fidelities, compared to Full-XY. 

Figure \ref{fig:fh} compares the fidelity for FH circuits with 10 and 20 qubits. At high noise levels, having multiple gate types offers significantly better executions, with average 1.45X, up to 1.7X improvement in fidelity compared to a single gate type, for the 20-qubit case. Across circuit sizes, the instruction set with multiple gate types (G7) outperforms the set with only single gate type (S2) consistently, but the benefits reduce as the hardware error rate improves. Multiple instruction types are most beneficial at hardware noise levels that are expected in the next 5-10 years.

To understand why reliability improves using multiple gate types, we studied two sources of improvements: 1) reduction in instruction counts and 2) noise adaptivity across instruction types. 
In Figures \ref{fig:rigetti_sim} and \ref{fig:sycamore_na}, the bars are annotated on top with the respective two-qubit instruction counts. Compared to S1-S7, the number of instructions required to implement applications reduces when multiple instruction types are available in R1-R5 and G1-G7.\footnote{Note that these reductions are less prominent than the heatmaps in Figure \ref{fig:fsim_heatmap} because the most expressive gate types were selected as the S1-S7 baselines.} For Rigetti, multiple-gate sets R1-R4 reduce the instruction count by up to 14\% compared to the single-instruction sets. When the SWAP gate is also added in R5, the reduction is on average 1.5X for QV, 2X for QAOA and 1.5X for QFT. Similarly, for Google, sets with multiple instruction types reduce the instruction count and G7 obtains average 1.9X reduction for QV, 1.6X for QAOA, 1.8X for QFT, and 1.3X for FH. In particular, the instruction counts of R5 match Full$_{\mathrm{XY}}$ for all the three applications and the counts of G7 are very close to  Full$_{\mathrm{fSim}}$. Hence, having a small number of instruction types is sufficient to obtain the same expressive power of a full continuous gate set for most applications.

To isolate the reliability benefits from gate count reductions from reductions obtained through noise adaptivity, in Figure \ref{fig:sycamore_qaoa_nu}, we repeated the experiment in Figure \ref{fig:sycamore_qaoa_na} but with the assumption that there is no noise variation across gate types. Compared to Figure \ref{fig:sycamore_qaoa_na}, G7 in Figure \ref{fig:sycamore_qaoa_nu} still reduces instruction counts and obtains higher cross entropy difference than the other discrete instruction sets. Importantly, it almost matches the performance of  Full$_{\mathrm{fSim}}$. In contrast, even though the gate count reductions of G1-G6 remain similar ($\sim10\%$), the application reliability improvements are lesser than the improvements in Figure \ref{fig:sycamore_qaoa_na}. Similar observations can be found for other benchmarks in Sycamore. This indicates that small improvements in gate counts are likely to have a low impact on application reliability in future systems where better gates and lower noise variability.
% and choosing one most expressive hardware gate for each application may be enough for systems that has no or small variations across gate types.
% G7 obtains higher heavy output probability than the preceding instruction sets and almost matches the performance of the fSimFull because it includes the SWAP gate.
However, in current and near-term systems, different gate types have significantly different error rates. \nuop~exploits these noise differences to choose the best gate types across qubits and therefore achieves more improvements with G1-G6 in Figure \ref{fig:sycamore_qaoa_na}. 

To study the importance of noise adaptivity, we compare the single-instruction sets S1-S4 with their combination G3 in Figure \ref{fig:sycamore_qft_na}. The two-qubit gates in S1-S4 have the same expressivity and same average gate error rates. However, S1-S4 have different performance because of fluctuations in error rates across the operating qubits. In comparison, G3 achieves much higher success rate than S1-S4 because exposing multiple instruction types allows noise-adaptivity across gate types and helps in mitigating the noise variations that arise when only type of gate is used. {\em Therefore, the fidelity improvements with multiple instruction types %in Figure~\ref{fig:rigetti_sim} and \ref{fig:sycamore_na} 
arise because of both improved expressivity and noise-adaptivity across gate types.}
%Current noise-adaptivie compilation strategies \cite{murali2019noise, tannu2019not} map the program qubits onto hardware qubits which have less noise.  

%Since, major hardware companies aim to double their quantum volume every year \cite{}, our results are very significant and show the importance of instruction set design for high fidelity executions. 
\begin{figure}[t]
    \centering
    \subfloat[]{
    \includegraphics[width=0.35\textwidth]{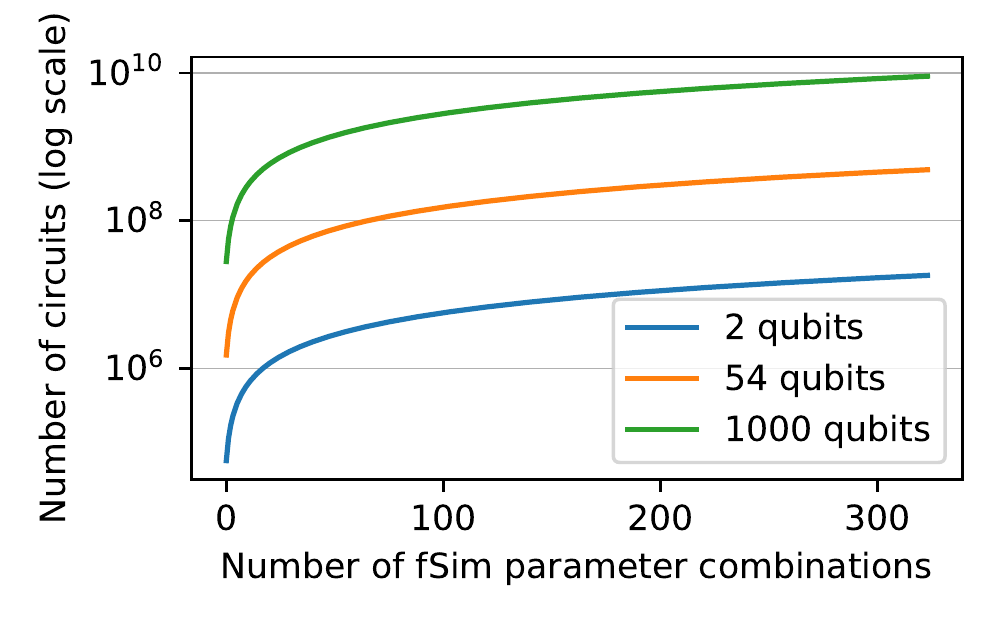}
    \label{fig:circuitnumber}
    }
    
    \subfloat[]{
    \includegraphics[width=0.35\textwidth]{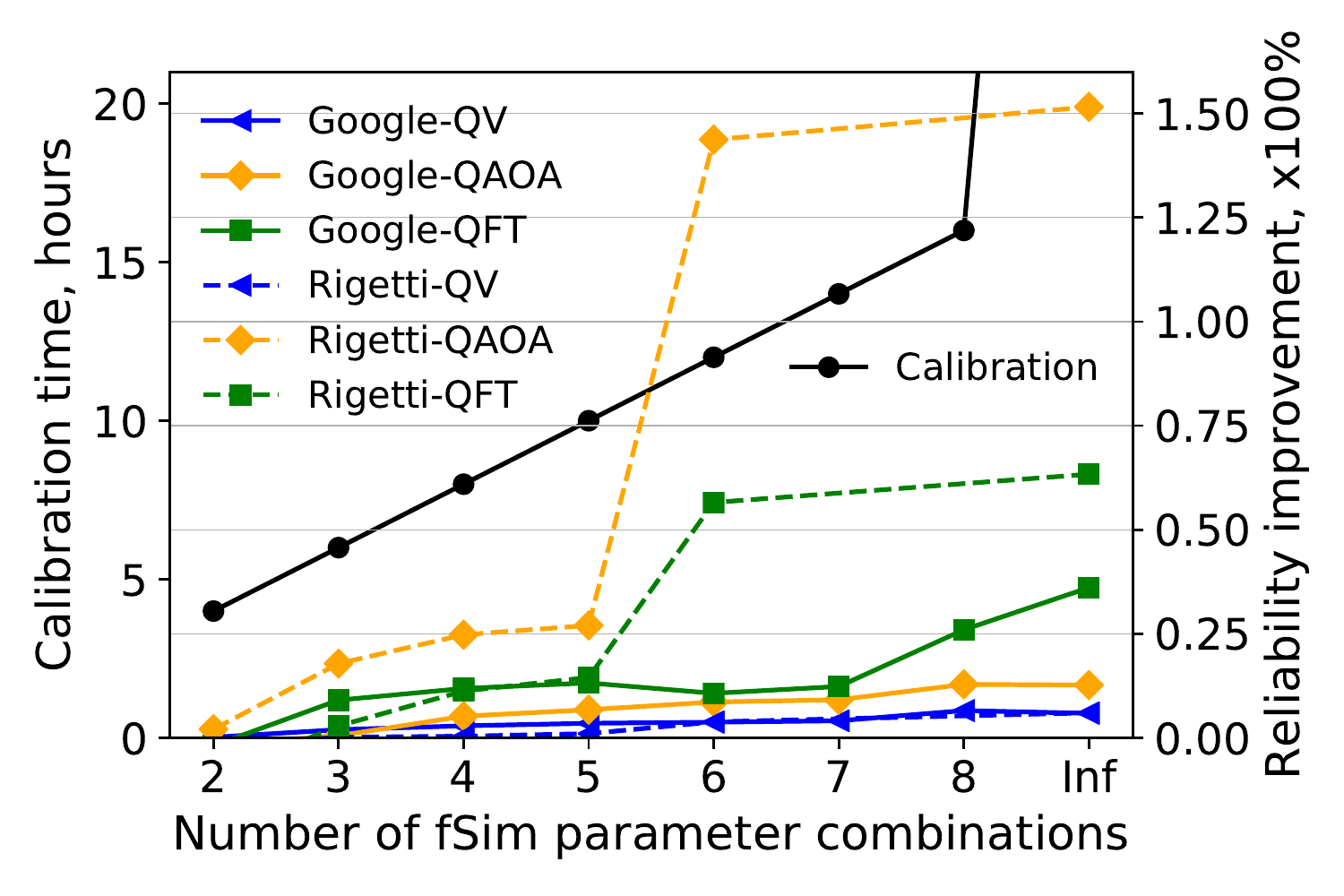}
    \label{fig:timecost}
    }
    \caption{Tradeoffs between calibration overhead and application performance. (a) The number of calibration and benchmarking circuits grows with the number of hardware gate types and the device size (\# qubits). (b) The calibration time cost (left y axis) and application performance improvement (right y axis) increase with the number of hardware gate types. 
    % Calibrating 8 types of gates may take more than 15 hours. 
    }
    \label{fig:calibration}
\end{figure}

\section{Calibration Overhead of Multiple Gate Types}
\label{sec:calibration}

In this section, we estimate the calibration overhead with multiple gate types by adopting the calibration model of the continuous fSim set in~\cite{google_continuous_gate}, where 525 gate types are calibrated on real hardware.
As mentioned previously, a fSim gate can be seen as the composite of a CPHASE gate and an iSWAP-like gate.
Therefore, the calibration of one single fSim gate type ($\mathrm{fSim(\theta, \phi)}$) on one qubit pair can be divided into several steps. The first step is to calibrate the CPHASE gates with rotation angle $\phi$ and angle $\pi$. The second step calibrates iSWAP-like gates with angle $0$ and angle $\pi/2$ for any preceding CPHASE angle $\phi$. Then one tunes up the iSWAP-like gate angle $\theta$ with CPHASE angle $\phi=\pi$.
Afterwards, the calibration data of these CPHASE and iSWAP-like gates are used to construct a pulse for the target $\mathrm{fSim(\theta, \phi)}$ gate and unitary tomography is performed to further adjust its pulse parameters.
Finally, the fidelity of this fSim gate is characterized by running a large number of cross entropy benchmarking circuits (1000 rounds). This is a conservative calibration model in which each gate type is calibrated individually on isolated qubits. In practice, more calibrations are required for minimizing errors due to pulse overlaps \cite{google_continuous_gate} and crosstalk. 

Figure~\ref{fig:circuitnumber} shows the number of calibration circuits for different number of gate types and system sizes. The number of circuits increases linearly with system size and the number of gate types. For a 54-qubit device, we require $\sim10^7$ circuits to calibrate 10 gate types. 
%The number of circuits becomes extremely large even for a 54-qubit device ($\sim10^7$ circuits for 10 gate types). 
For a 1000-qubit system, nearly a billion circuits are required even for a small number of gate types. 
%This in turn increases the time required for calibration.
Comparing with current systems, calibrating two qubits in \cite{arute2019quantum} takes up to four hours, including the calibration of electronics, qubit frequencies, single-qubit rotations, and a single two-qubit gate type. Even if we conservatively assume that a single two-qubit gate type takes $\sim$2 hours, the linear scaling of calibration time with gate types and system sizes makes it practically intractable to calibrate a full gate family.

Figure~\ref{fig:timecost} shows the tradeoffs between calibration time and application reliability (data from Figures \ref{fig:rigetti_sim} and \ref{fig:sycamore_na}).
When more gates are available in the hardware, the average application reliability improves, but with diminishing returns after five gate types. Interestingly, the gate sets that have SWAP gates (R5 and G7) offer reliability comparable to the continuous sets.
In practice, the continuous instruction set will likely have lower average gate fidelity than the discrete sets due to calibration difficulties and pulse overlaps~\cite{eth_two_qubit_gate}. The reliability improvements resulting from gate reduction may be compromised by the increased average gate error rates. To study this, we performed simulations using Full$_{\mathrm{fSim}}$ with different increasing factors in the average error rates as shown in Figure~\ref{fig:sycamore_qv_na}-\ref{fig:sycamore_qft_na}. Compared to discrete multi-gate sets G1-G7 with average error rates 0.62\%, the advantages of Full$_{\mathrm{fSim}}$ disappear when its average error rates increases to 1.5X, 2X, 2.5X for QV, QAOA, and QFT benchmarks respectively. 

Since calibration and benchmarking overheads increase with the number of gate types, having a small number of expressive two-qubit gates is beneficial and sufficient to obtain high reliability.
\emph{To summarize, we recommend that QC systems implement 4-8 different two-qubit hardware gates that are expressive for most applications to obtain a sweet spot of high-fidelity application executions and low device calibration overheads. In addition, having hardware SWAP gates for connectivity-constrained devices or other application-specific gates can greatly improve reliability.}

\section{Conclusions}
\label{sec:conclusion}
%Early NISQ systems implemented instruction sets with a single two-qubit gate type. 
Most QC instruction sets have a single type of two-qubit gate. This design choice sacrifices application expressivity, but simplifies calibration. On the other hand, vendors have proposed continuous gate families which maximize application expressivity, but are very difficult to calibrate especially as systems scale up. These tradeoffs are reminiscent of the instruction count vs. hardware complexity tradeoffs of RISC and CISC designs in classical computer architecture. 
Our work used an application-based simulation analysis of instruction sets, with realistic device and calibration models, to identify a small instruction set that is both {\em efficient for application decompositions and easy to calibrate}. 
Our proposed instruction sets R5 for Rigetti and G7 for Google have only 4-8 two-qubit gate types, but offer near-optimal application-expressivity and fidelity compared to a continuous instruction set, while reducing calibration overhead by two orders of magnitude. NuOp is available open-source at \url{https://github.com/prakashmurali/NuOp}.

%Our work is the first to study the application expressivity vs. calibration tradeoff for QC instruction sets. 
With several vendors aiming to improve QC reliability by offering better gate types, our work is poised to guide instruction set design choices through the NISQ-regime. 

\section{Acknowledgements}
 We thank Eric Peterson (Rigetti), Nathan Lacroix (ETH Zurich), Zhang Jiang (Google) and Tom O'Brien (Google) for discussions on compilation and instruction set design. This work is funded in part by Enabling Practical-scale Quantum Computation (EPiQC), an NSF Expedition in Computing, grant 1730082. P.M was also supported by an IBM PhD Fellowship. D.B and L.L acknowledge support from the EPSRC Prosperity Partnership in Quantum Software for Modelling and Simulation, grant EP/S005021/1.
%%%%%%% -- PAPER CONTENT ENDS -- %%%%%%%%

%%%%%%%%% -- BIB STYLE AND FILE -- %%%%%%%%
\bibliographystyle{IEEEtran}
\bibliography{references}
\end{document}